\documentclass[sn-mathphys-num]{sn-jnl}

\usepackage{graphicx}%
\usepackage{multirow}%
\usepackage{amsmath,amssymb,amsfonts}%
\usepackage{amsthm}%
\usepackage{mathrsfs}%
\usepackage[title]{appendix}%
\usepackage{xcolor}%
\usepackage{textcomp}%
\usepackage{manyfoot}%
\usepackage{booktabs}%
\usepackage{algorithm}%
\usepackage{algorithmicx}%
\usepackage{algpseudocode}%
\usepackage{listings}%
\usepackage{bm}
\usepackage{pifont}

\newcommand{\ignore}[1]{}
\usepackage{xspace}

\newcommand{\ie}{\emph{i.e.,}\xspace}

\newcommand{\eg}{\emph{e.g.,}\xspace}

\newcommand{\wrt}{w.r.t.\xspace}
\newcommand{\ourmodel}{CKD-MDSR}
\newcommand{\blue}{\textcolor{blue}}

\theoremstyle{thmstyleone}%
\theoremstyle{thmstyletwo}%

\theoremstyle{thmstylethree}%

\begin{document}

\title[Article Title]{Curriculum-scheduled Knowledge Distillation from Multiple Pre-trained Teachers for Multi-domain Sequential Recommendation}




\author*[1]{\fnm{Wenqi} \sur{Sun}}\email{wenqisun@ruc.edu.cn}

\author[2]{\fnm{Ruobing} \sur{Xie}}\email{xrbsnowing@163.com}

\author[1]{\fnm{Junjie} \sur{Zhang}}

\author[1]{\fnm{Wayne Xin} \sur{Zhao}}\email{batmanfly@gmail.com}

\author[2]{\fnm{Leyu} \sur{Lin}}

\author[1]{\fnm{Ji-Rong} \sur{Wen}}

\affil*[1]{\orgdiv{Gaoling School of Artificial Intelligence}, \orgname{Renmin University of China}, \orgaddress{\postcode{100872}, \country{China}}}

\affil[2]{\orgname{Tencent Inc.}, \orgaddress{\city{Beijing}, \postcode{100193}, \country{China}}}



\abstract{Pre-trained recommendation models (PRMs) have received increasing interest recently. However, their intrinsically heterogeneous model structure, huge model size and computation cost hinder their adoptions in practical recommender systems.
Hence, it is highly essential to explore how to use different pre-trained recommendation models efficiently in real-world systems.
In this paper, we propose a novel curriculum-scheduled knowledge distillation from multiple pre-trained teachers for multi-domain sequential recommendation,
called \ourmodel{}, which takes full advantages of different PRMs as multiple teacher models to boost a small student recommendation model, integrating the knowledge across multiple domains from PRMs.
Specifically, \ourmodel{} first adopts curriculum-scheduled user behavior sequence sampling and distills informative knowledge jointly from the representative PRMs such as UniSRec and Recformer.
Then, the knowledge from the above PRMs are selectively integrated into the student model in consideration of their confidence and consistency.
Finally, we verify the proposed method on multi-domain sequential recommendation and further demonstrate its universality with multiple types of student models, including feature interaction and graph based recommendation models.
Extensive experiments on five real-world datasets demonstrate the effectiveness and efficiency of \ourmodel{}, which can be viewed as an efficient shortcut using PRMs in real-world systems.
Our code is available at~\url{https://github.com/RUCAIBox/CKD-MDSR}.
}

\keywords{Multi-domain sequential recommendation, Pre-trained recommendation model, Curriculum knowledge distillation}



\maketitle

\section{Introduction}
Nowadays, pre-trained recommendation models (PRMs) (\eg UniSRec \cite{UniSRec2022} and Recformer \cite{Recformer2023}) have garnered widespread attention in the research community of recommender systems (RS).
This interest stems from their ability to enhance recommendation performance by leveraging extensive interaction data across various domains.
Recent efforts on PRMs have made significant progresses \cite{PeterRec2020,qiu2021U-BERT,mao2023unitrec,MISSRec2023}.
The mainstream pre-trained models for multi-domain sequential recommendation can be roughly grouped into three categories.
The first category is based on representing items as semantic latent vectors, which derive from item texts via the pre-trained language models (PLMs), and then leveraging the conventional sequential recommendation (SR) models like SASRec \cite{SASRec} for behavior modeling (\eg UniSRec~\cite{UniSRec2022}).
The second category resorts to directly converting items as plain texts and then encoding the text sequence with pre-trained language models (\eg Recformer~\cite{Recformer2023}).
The final category relies on leveraging the multimodal side information of items via the pre-trained vision-language models and then learning the item representations and sequential behavior patterns (\eg UniM$^2$Rec~\cite{UniM2Rec2023}).
Through pre-training, substantial informative knowledge from multiple domains can be integrated into PRMs.
This integration has been shown to yield promising results in downstream tasks, such as cold-start and multi-domain sequential recommendations.

\begin{figure}[ht!]
	\centering
	\includegraphics[width=0.9\linewidth]{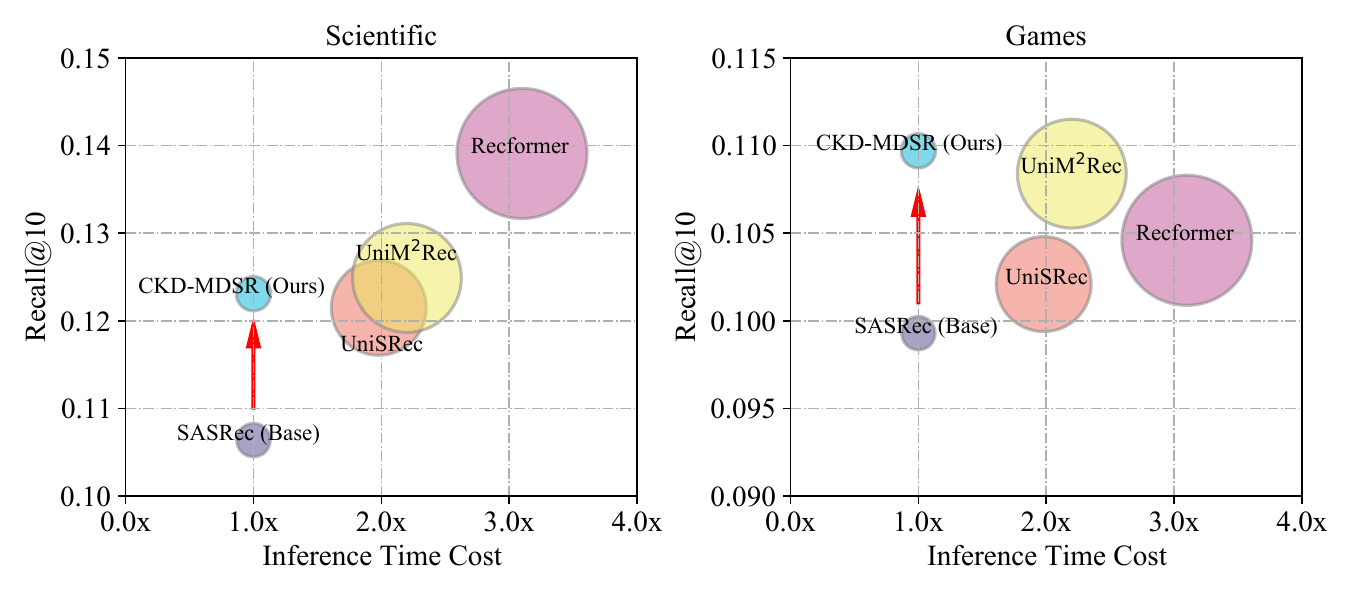}
	\caption{Model comparisons \wrt the trade-off among inference time cost ($x$ axis), performance ($y$ axis, the geometric center of these circles), and memory cost (size of circles) on two datasets. Our proposed \ourmodel{} achieves good accuracy with the lowest online computation and memory costs.}
	\label{fig:efficiency}
\vspace{-0.3cm}
\end{figure}

Despite the progress, existing pre-trained recommendation models still struggle with three issues in their practical usage:
(\textbf{i}) PRMs have huge model sizes and computation costs, making it hard to directly deploy PRMs in the low-latency and compute-intensive online systems \cite{lightweight2021}. 
The efficiency regarding computational costs and complexity is a critical concern in practical recommender systems, particularly in the current era characterized by a significant shortage of computing power.
(\textbf{ii}) The architectures of existing PRMs significantly differ from those of practical recommender systems, hindering seamless integration within online systems \cite{UniSRec2022,Recformer2023}. Practical recommender systems typically incorporate various sequential recommendation (SR) models in different domains (\eg SASRec \cite{SASRec} and FMLP-Rec \cite{FMLP-Rec2022}), which cannot be fully and smoothly addressed by these PRMs.
(\textbf{iii}) Different PRMs exhibit heterogeneous features and outputs, often demonstrating distinct advantages in different domains.
No dominant PRMs (even compared to conventional SR models) excel across all recommendation tasks and scenarios, unlike the unparalleled capabilities of GPT-4 in natural language processing (NLP). Therefore, relying solely on a single PRM may result in a loss of potential enhancements that could be achieved by integrating heterogeneous PRMs.
In light of these challenges, it is essential to develop an \emph{effective}, \emph{low-cost}, \emph{model-agnostic}, and \emph{easy-to-deploy} framework that harnesses the distinct strengths of heterogeneous PRMs to enhance practical recommendation systems.
\emph{Knowledge distillation (KD)}, which facilitates the transfer of informative knowledge from teacher model(s) to smaller student model(s) with smaller model size, presents a promising approach for addressing the aforementioned challenges \cite{hinton2015distilling,UnbiasedKD2023}.

In this work, we propose a novel \textbf{curriculum-scheduled knowledge distillation from multiple pre-trained teachers for multi-domain sequential recommendation (called \ourmodel{})}, which could selectively integrate the informative knowledge distilled from different PRMs into a student model.
Specifically, \ourmodel~distills the knowledge from three representative types of PRMs including UniSRec \cite{UniSRec2022}, Recformer \cite{Recformer2023}, and UniM$^2$Rec \cite{UniM2Rec2023} to maximize the diversity of knowledge sources and enhance recommendation performance.
These PRMs vary in terms of features, outputs, training losses, item representations, and sequential modeling approaches, making their joint integration challenging.
Besides, PRMs may generate predictions that contain potential noisy knowledge.
To address these issues, our proposed method incorporates curriculum-scheduled sequence sampling and distills the scores of in-batch negative samples from PRMs to facilitate knowledge transfer, effectively leveraging the commonalities among different PRMs.
Additionally, we devise an instance-level scoring strategy to refine knowledge based on consistency across PRMs.
Finally, We validate the proposed method on multi-domain sequential recommendation and further demonstrate its universality with multiple types of student models (encompassing feature interaction- and graph-based models), with the aim of simulating real-world recommender systems.
In experiments, we conduct extensive evaluations on five real-world datasets, where \ourmodel{} achieves consistent improvements over its corresponding base model. Besides, we conduct the ablation study, universality analyses and model analyses for in-depth understanding.
Our main contributions are summarized as follows:

\begin{itemize}
    \item We propose a novel curriculum-scheduled knowledge distillation from multiple pre-trained teachers for multi-domain sequential recommendation. To best of our knowledge, \ourmodel{} is the first KD framework oriented to multiple PRMs, which could integrate the knowledge from heterogeneous PRMs across various domains efficiently.
    \item We propose an instance-level scoring strategy, which selects informative knowledge from heterogeneous PRMs according to their confidence and consistency.
    \item Extensive experiments conducted on five real-world datasets demonstrate the effectiveness, efficiency and universality of \ourmodel{}. It could be regarded as an efficient shortcut in practical recommender systems.
\end{itemize}

\section{Related Work}
Our work is closely related to study on pre-trained recommendation models, multi-domain sequential recommendation and knowledge distillation for recommendation.

\subsection{Pre-trained Recommendation Models}\label{sec:prm}
With the rapid advancement of pre-trained language models (PLMs), adapting PLMs for recommender systems (RS) presents a promising opportunity for performance enhancement.
Consequently, an increasing number of RS researchers are focusing on pre-trained recommendation models, often referred to as recommendation foundation models~\cite{huaweirec2023,LLMRec2024,WWWJ-sequential-recommendation}.
Currently, the mainstream pre-trained approaches in multi-domain sequential recommendation can be broadly classified into three categories.
(\textbf{i}) The first category of methods relies on learning item embeddings from item texts to serve as a bridge across multiple domains.
Specifically, UniSRec \cite{UniSRec2022} employs a PLM with domain adaptors to construct universal item representations from texts and leverages conventional sequential models (\eg SASRec \cite{SASRec}) for sequential behavior modeling.
(\textbf{ii}) The second category of approaches involves converting behavior sequences into text sequences and applying PLMs directly for sequential behavior modeling. As an example, Recformer \cite{Recformer2023} treats items as plain texts and employs a behavior-tuned PLM (\ie LongFormer \cite{beltagy2020longformer}) to encode the sequence of historical behaviors, predicting the next item accordingly.
(\textbf{iii}) The third category of methods integrates pre-trained textual and visual models to enhance item representations.
To be specific, UniM$^2$Rec \cite{UniM2Rec2023} is a representative multi-modal PRM that utilizes diverse multi-modal information, such as texts and images, as supplementary bridges across various downstream domains.
It is worth noting that there are significant differences in the focus of each pre-trained recommendation model, distinguishing them from models in NLP.
Furthermore, given that recommendation models are widely deployed in various online services, their efficiency is emphasized more in the field of RS than in NLP.
Hence, it is both crucial and necessary to investigate the efficiency of PRMs in practical recommendation scenarios.

\subsection{Multi-domain Sequential Recommendation}
Multi-domain sequential recommendation (MDSR) aims to improve the performance of sequential recommenders in multiple target domains leveraging knowledge transferred from source domains~\cite{STAR2021,MDSR-2024,WWWJ-Cross-domain}.
The key aspects of multi-domain sequential recommendation involve the effective integration of interaction data from multiple source domains and the transfer of knowledge from source domains to target domains~\cite{AFT2021,ADIN2022}.
MDSR approaches often encompass the user interactions and item modality information from source domains to train SR models.
Most existing multi-domain sequential recommenders generally leverage the item multimodal side information to represent them as semantic items, thereby enabling effective transfer across different domains.
Specifically, $\pi$-Net~\cite{PINet2019} and PSJNet~\cite{PSJNet2023} employ RNN to generate user-specific representations,
which emphasize the sequential dependencies but fail to depict transitions among associated entities.
CCDR~\cite{CCDR2022} proposes two intra-domain and inter-domain contrastive learning tasks to enhance cross-domain recommendation, where the aligned users, tags, words, and categories are functioned as the semantic bridges across different domains.
UniSRec~\cite{UniSRec2022} utilizes item texts processed by pre-trained language model to get the universal item representation as anchors in all domains.
MISSRec~\cite{MISSRec2023} proposes an interest-aware multimodal sequential recommendation method that utilizes multimodal item contents and multi-domain user interactions.
In this work, we mainly focus on effectively and efficiently integrating knowledge from heterogeneous PRMs into a student recommendation model for multi-domain sequential recommendation.

\subsection{Knowledge Distillation for Recommendation}
Knowledge distillation (KD) \cite{hinton2015distilling}, which involves training a lightweight student model by extracting knowledge from a cumbersome teacher model, is widely employed for learning efficient PLMs in NLP~\cite{DistillStep2023}.
However, in contrast to the various PLMs in NLP, which typically yield similar generated results, the next-item predictions from diverse pre-trained recommendation models exhibit significant differences~\cite{DualDistill2021,PnKD2022,UnbiasedKD2023}.
Thus, it is essential to explore the most suitable methods for efficient recommendation, instead of simply borrowing existing methods from NLP.
To address the efficiency issue in PRMs, an intuitive approach is to distill knowledge from PRMs and integrate it into a compact downstream recommendation model via knowledge distillation.
In the RS field, many efforts have been made to apply knowledge distillation to enhance recommendation~\cite{AdversarialDistillation2019,RankingDistillation2018,BidirectionalKD2021,CFKD2019,KDRS2020}.
Specifically,  HetComp~\cite{HetComp2023} proposes leveraging dynamic knowledge construction to provide progressive knowledge transfer, enabling the gradual assimilation of fine-grained ranking information.
UnKD~\cite{UnbiasedKD2023} proposes a stratified distillation strategy for recommendation.
Different from existing works~\cite{HetComp2023,UnbiasedKD2023,EMKD2023}, our proposed method focuses on a KD framework that integrates knowledge from heterogeneous PRMs across multiple domains.

\section{Methodology}
In this section, we present the details of our proposed curriculum-scheduled knowledge distillation framework, utilizing heterogeneous pre-trained teachers for multi-domain sequential recommendation.
We first introduce the motivation of our method and the problem formulation, and then describe the structure of \ourmodel{} and model training in detail.

\subsection{Motivation}
Recently, PLMs have achieved remarkable successes in various NLP tasks, and researchers in the RS field attempt to bring the potential power of PLMs to recommendation \cite{P5-2022,Recformer2023}.
The additional semantic knowledge and powerful sequential modeling ability of PLMs do help to alleviate the issue of data sparsity in recommendation, especially in zero-shot scenarios \cite{Recformer2023}.
However, heterogeneous PRMs possess distinct characteristics tailored for diverse scenarios (\eg the multimodal modeling in UniM$^2$Rec \cite{UniM2Rec2023} offers distinct advantages on micro-video scenarios)
and demonstrate their advantages across different domains (as shown in Table \ref{tab:performance}).
Furthermore, even PRMs tuned on user behavioral information cannot achieve the dominant performance over conventional SR models in all domains \cite{hou2023large}.
No dominant PRMs excel across all recommendation tasks and scenarios, unlike pre-trained language models in NLP.
Besides, the efficiency issue \wrt computational costs and complexity is a key obstacle hindering the practical deployment of PRMs in the RS field.
To this end, our goal is to extract informative knowledge from heterogeneous PRMs and integrate it into a student model, enabling it to be highly efficient and proficient at effectively absorbing knowledge from different PRMs.

\begin{figure}[t]
	\centering
	\includegraphics[width=0.99\linewidth]{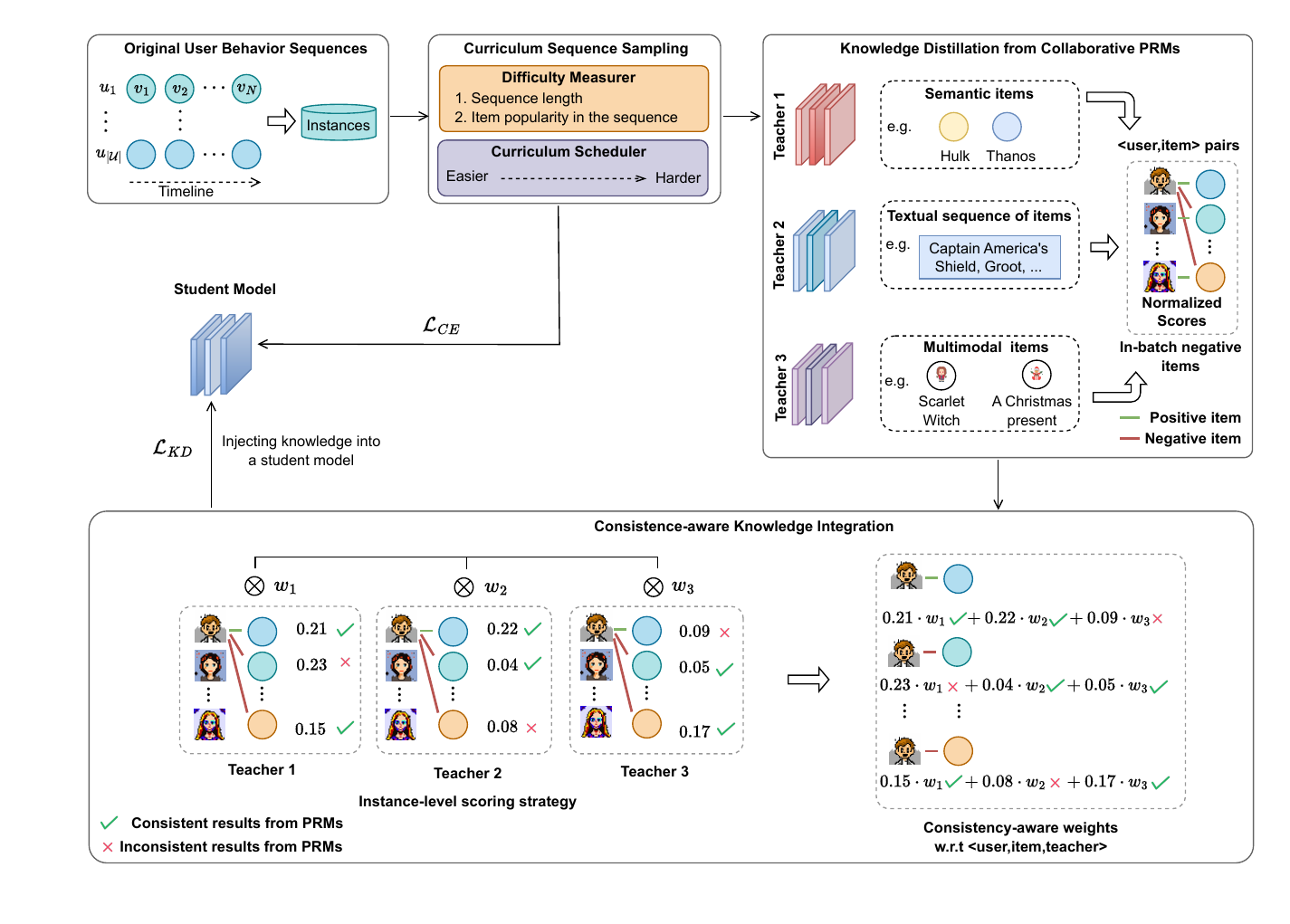}
	\caption{An overview of our proposed \ourmodel{}.}
	\label{fig:model}
\end{figure}

\subsection{Problem formulation}
The goal of multi-domain sequential recommendation (MDSR) is to improve the performance of sequential recommenders in multiple target domains leveraging knowledge transferred from source domains.
Now, we start with a brief description of typical MDSR.
Let $\mathcal{U}$ and $\mathcal{V}$ be a set of users and items in a target domain, respectively.
Given a user $u$, her/his interactions sorted by the timestamp is an item sequence $S^u=\{v^u_1,v^u_2,\ldots,v^u_N\}$ in the target domain.
MDSR aims to predict next item $v^u_{(T+1)}$ that $u$ would probably interact with based on her/his historical interactions, while leveraging knowledge from source domains to enhance predictions in the target domain.
Note that MDSR usually pre-trains the item multimodal side information and user interactions in source domains to extract transferable knowledge, without requiring overlapping users or items between the source and target domains.

\subsection{Approach Overview}
Our proposed KD framework consists of three major modules, namely curriculum sequence sampling, knowledge distillation from multiple PRMs, and consistency-aware knowledge integration.
Specifically,
(\textbf{i}) to facilitate the model learning, we adopt an easy-to-hard user sequence sampling strategy in consideration of the steep learning difficulty of user behavior sequences across various domains and the noisy knowledge in the KD process.
(\textbf{ii}) To integrate the capabilities of heterogeneous PRMs across various domains, we distill the knowledge from different PRMs via the in-batch negative sampling strategy.
(\textbf{iii}) To address the efficiency issues in PRMs, we integrate the knowledge into a student recommendation model.
In addition, to address inconsistencies and potential noise in outputs across different PRMs, we propose a instance-level scoring strategy and devise consistency-aware weights, which selects informative knowledge from heterogeneous PRMs according to their confidence and consistency, thereby extracting valuable knowledge.
The overall framework of the proposed \ourmodel{}~is illustrated in Fig.~\ref{fig:model}.

\subsection{Curriculum-scheduled User Behavior Sequence Sampling}
Considering the steep learning difficulty of user behavior sequences across various domains and the presence of noisy user preferences in the KD process,
we propose an easy-to-hard user behavior sequence sampling strategy,
which is inspired by principles of curriculum learning~\cite{CL_rec}.
Curriculum learning involves training a model on progressively more difficult data, mimicking the human learning process, which has been successfully applied across various fields~\cite{CL_DNN, CL_NLP, CL_IR, CL_rec} to enhance learning effectiveness.
At the core of curriculum learning is the design of a difficulty measurer, which assesses the learning difficulty of each sequence sample, and a curriculum scheduler, which determines the dataset partitioning and scheduling throughout the training process.
Hence, we detail our curriculum-scheduled user sequence sampling strategy from these two aspects.

\subsubsection{Difficulty Measurer}

An appropriate difficulty measurer is crucial for the effectiveness of curriculum learning.
However, to the best of our knowledge, it remains unclear what difficulty measurers are effective for multi-domain sequential recommendation.
In this work, we propose to leverage the \textit{sequence length} and \textit{item popularity}-based difficulty measurer for user behavior sequence samples.
Specifically, for a sequence sample $S^u=\{v^u_1,v^u_2,\ldots,v^u_N\}$ with length $N$, we comprehensively consider both the sequence length and the popularity of each item within the sequence, defining the sequence sample difficulty (SSL) as follows:

\begin{equation}
    \text{SSL}(S^u)=\frac{N}{N_{max}}-\alpha\frac{\sum_{i=1}^N\text{pop}(v^u_i)}{N},
\end{equation}
where $\alpha>0$ is a hyperparameter,
and $N_{max}$ denotes the maximum sequence length in the corresponding dataset,
and pop() denotes the item popularity, which we derive following the the previous work~\cite{UnbiasedKD2023}.
Intuitively, sequences with longer lengths and less popular items (\ie larger SSL) are harder for learning their corresponding user preferences.
Thus, SSL serves as an appropriate difficulty measurer for sequence samples in multi-domain sequential recommendation.
We further stipulate that a larger SSL corresponds to a harder sequence sample.

\begin{algorithm}[!ht]
	\renewcommand{\algorithmicrequire}{\textbf{Input:}}
	\renewcommand{\algorithmicensure}{\textbf{Output:}}
\caption{The curriculum-scheduled user behavior sequence sampling strategy.}\label{alg:curriculum}

\begin{algorithmic}[1]
\Require training sequence set $\mathcal{D}$.
\Ensure the sorted training sequences, the learned sequential recommender.
\State Sorting $\mathcal{D}$ according to their SSL
\State Dividing $\mathcal{D}$ into $B$ buckets
\For{$b = 1,\ldots,B$}
    \State Training on the buckets $B_{1:r}$
\EndFor
\Repeat
    \State Training on full data that randomly samples from $\mathcal{D}$
\Until converged
\end{algorithmic}
\end{algorithm}

\subsubsection{Curriculum Scheduler}
Considering all training sequence samples $\mathcal{D} = \{S^{u_1}, S^{u_2}, ..., S^{u_{|\mathcal{U}|}}\}$ that are sorted in ascending order of difficulty,
these sorted sequences are then evenly divided into $B$ buckets, ranging from the easiest to the hardest, where $B$ is a hyperparameter.
These buckets can be viewed as categorizing the samples into a learning curriculum on different difficulty levels.
Then, we train the sequence samples in multiple stages, progressing through each bucket one by one with higher difficulty for optimization.
At stage $t_r$, we use the buckets $B_{1:r}$ as the training data.
Once all buckets have been utilized, the model returns to the standard training mode, where samples are drawn randomly from $\mathcal{D}$, and the remaining predefined training epochs are completed until convergence.
Through this approach, \ourmodel{} effectively learns curriculum-scheduled user behavior sequences, starting with easier sequences and gradually progressing to harder ones.
The training process is summarized in Algorithm~\ref{alg:curriculum}.

\subsection{Knowledge Distillation from Multiple PRMs}

In this part, we present the model training and related discussions of our proposed approach.

\subsubsection{Pre-trained Recommendation Models in \ourmodel{}}

Pre-trained recommendation models can incorporate extensive knowledge from source domains and transfer this knowledge to target domains, thereby enhancing recommendation performance.
Without loss of generality, we select three representative models (\ie UniSRec, Recformer, and UniM$^2$Rec) from three categories (shown in Section~\ref{sec:prm}) as teacher models.
The summary of characteristics for these PRMs is presented in Table~\ref{tab:summary}. It is evident that the input features, item representations, and training losses vary significantly across these PRMs, making the ensemble of these models and the extraction of informative knowledge challenging.
It is worth noting that the proposed KD framework is universal, facilitating the flexible replacement of teacher models.

\begin{table}[!ht]
    \centering
    \caption{Characteristics of three representative PRM serving as teacher models.}
    \label{tab:summary}
    \begin{tabular*}{\textwidth}{@{\extracolsep\fill}l|c|c|c|c|c}
        \toprule
         \multicolumn{1}{c|}{\multirow{2}{*}{\textbf{PRM}}} & 
         \multicolumn{3}{c|}{\textbf{Input Feature}} & 
         \multirow{2}{*}{\textbf{Item Representation}} & 
         \multirow{2}{*}{\textbf{Training Loss}} \\ \cmidrule{2-4}
           & \textbf{ID} & \textbf{Text} & \textbf{Image} & & \\
        \midrule
        UniSRec~\cite{UniSRec2022} & \textcolor{teal}{\ding{51}} & \textcolor{teal}{\ding{51}} & \textcolor{purple}{\ding{55}} & 
        Textual (+ID) Emb. &
        
        Cross-Entropy \\ 
        
        Recformer~\cite{Recformer2023} & \textcolor{purple}{\ding{55}} & \textcolor{teal}{\ding{51}} & \textcolor{purple}{\ding{55}} & 
        Plain Texts &
        MLM + IIC\footnotemark[1] \\
        
        UniM$^2$Rec~\cite{UniM2Rec2023} & \textcolor{teal}{\ding{51}} & \textcolor{teal}{\ding{51}} & \textcolor{teal}{\ding{51}} & 
        Textual/Visual (+ID) Emb. &
        Cross-Entropy \\
        \bottomrule
    \end{tabular*}
    \footnotetext[1]{``MLM'' and ``IIC'' denote masked language modeling and item-item contrastive task respectively.}
\end{table}

\subsubsection{In-batch Negative Sampling}

Given the varying structures and outputs of PRMs, \ourmodel~distills the normalized user-item scores from the in-batch negative samples for each user during training. This approach establishes a common invariant across different PRMs in the knowledge distillation process, enabling distillation without the need to consider specific model structures and outputs.
We propose to leverage in-batch negative samples for distillation instead of the entire item corpus, since (\textbf{i}) the entire corpus is too large, compromising the efficiency of distillation, while the size of in-batch negative samples is manageable, and (\textbf{ii}) in-batch negative samples are generally of higher quality and often represent popular items, which are more confused to predict and therefore require additional learning.
Conversely, distilling scores from all items could introduce irrelevant knowledge and noise (as PRMs are less confident with long-tail items), potentially degrading the performance of knowledge distillation.

Consider a batch of $T$ training instances, where each positive instance is a user-item pair.
We encode them into the representation set $\mathbb{P}=\{(\bm{u}_{1}, \bm{e}_{1}), \ldots, (\bm{u}_{T}, \bm{e}_{T})\}$, where $\bm{u}_i$ represents the representation of user $i$ and $\bm{e}_j$ denotes the representation of item $j$.
From this, we derive the in-batch item set $\mathbb{E}=\{\bm{e}_{1}, \ldots, \bm{e}_{T}\}$, the negative samples of user $i$ are all items of $\mathbb{E}$ except $\bm{e}_i$.
Specifically, the distillation score $s_{\langle i,j,k\rangle}$ for a user-item pair $\langle\bm{u}_i,\bm{e}_j\rangle$ from teacher $k$ is designed as follows:
\begin{equation}
    \text{s}_{\langle i,j,k\rangle}=\frac{\exp{\left(\bm{u}_i\cdot\bm{e}_j/\tau\right)}}
    {\sum_{e_{j^{\prime}}} \exp{\left(\bm{u}_i\cdot\bm{e}_{j^{\prime}}/\tau\right)}},
    \quad \bm{e}_{j^{\prime}} \in \mathbb{E},
\end{equation}
where $\tau>0$ is the temperature hyperparameter.

\subsection{Consistency-aware Knowledge Integration}

The scores distilled from different PRMs on the same user-item sample may vary in correctness and confidence, since no dominant PRMs can handle all unexposed $\langle \bm{u}_i,\bm{e}_j\rangle$ for predictions.
Thus, we allocate a weight $w_{\langle i,j,k\rangle} \in [0,1]$ for the teacher $k$ and the weight $w_{\langle i,j,k\rangle}$ mainly depend on the empirical strengths of the $k$-th PRM in the specific target domains.
the weights assigned to each teacher comply with:
\begin{equation}
    \sum_k w_{\langle i,j,k\rangle} = 1.
\end{equation}

Then, to alleviate noises generated by PRMs within the distillation scores, we propose the instance-level scoring strategy for learning from multiple PRMs and devise consistency-aware weights.

\subsubsection{Instance-level Scoring Strategy}
For the scores of each distilled instance, we utilize dynamic weights to primarily select knowledge with predominantly validated information and high confidence. To be specific, given a user and the in-batch samples denoted as $\bm{u}_{i}$ and $\{\bm{e}_{1}, \ldots, \bm{e}_{T}\}$, the scores from different PRMs exhibit different data distributions.
To mitigate this differences, we first normalize these scores of each PRM to make them relatively comparable, obtaining the normalized scores $\widetilde{\text{s}}_{\langle i,j,k\rangle}$ from $\text{s}_{\langle i,j,k\rangle}$.
Then, we utilize the \textbf{mean-square error} (MSE) to measure the consistency and differences between different PRMs. Precisely, we obtain the MSE value between all $\widetilde{\text{s}}_{\langle i,j,k\rangle}$ of any two teachers, and get the total sum of MSE scores (regarded as the confidence score $\text{conf}_{\langle i,j,k\rangle}$) for each teacher $k$ of the user $\bm{u}_i$ are given as follows:
\begin{equation}
\text{conf}_{\langle i,j, k\rangle}=-\sum_{k' \neq k} (\widetilde{\text{s}}_{\langle i,j,k\rangle}-\widetilde{\text{s}}_{\langle i,j,k'\rangle})^2.
\end{equation}
We can find that a lower confidence score indicates that the corresponding PRM $k$ (\ie a teacher model) is not consistent with other PRMs for the current user $\bm{u}_i$, which implies that the score $\text{s}_{\langle i,j,k\rangle}$ is not convinced.

\subsubsection{Consistency-aware Weights}

If all three confidence scores $\text{conf}_{\langle i,k\rangle}$ are less than a threshold $\epsilon$, we retain all scores $s_{\langle i,j,k\rangle}$ for distillation with the default weights $w_{\langle i,j,k\rangle}$. Otherwise, we exclude the distillation scores from the teacher with the highest $\text{conf}_{\langle i,k\rangle}$ (\ie the least confident PRM), redistributing its weight $w_{\langle i,j,k\rangle}$ equally to the other teachers.
Through this instance-level scoring strategy, we get a new set of distillation weights as $\hat{w}_{\langle i,j,k\rangle}$. The final weighted score $\hat{\text{s}}_{\langle i,j\rangle}$ of each $\langle \bm{u}_i,\bm{e}_j\rangle$ in the multi-teacher distillation is formalized as follows:
\begin{equation}
\hat{\text{s}}_{\langle i,j\rangle}=\sum_k \hat{w}_{\langle i,j,k\rangle} \widetilde{\text{s}}_{\langle i,j,k\rangle},
\end{equation}
where $\widetilde{\text{s}}_{\langle i,j,k\rangle}$ is the normalized score.

Finally, the knowledge distilled from different PRMs $\text{s}_{\langle i,j,k\rangle}$ are integrated into a student model.
Without loss of generality, we select the typical SASRec~\cite{SASRec} as our student model in main evaluations.
Note that our proposed \ourmodel{}~is universal with various types of downstream student models, encompassing sequential recommendation, feature interaction- and graph-based models.

\subsection{Model Training and Discussion}

In this part, we present the model training and comprehensive discussions of our proposed approach.

\subsubsection{Model Training}
The training of the student model comprises two principal components: the original supervised training and the knowledge distillation from the teacher models.
For the original supervised training, we follow the training of the base model (\eg SASRec) and adopt the \emph{cross-entropy loss} as follows:
\begin{equation}
    \mathcal{L}_{CE}=- \sum_{\langle \bm{u}_i,\bm{e}_j\rangle \in \mathbb{E}} \log{\frac{\exp{(\langle\boldsymbol{u}_i,\bm{e}_j\rangle)}}{\sum_{\langle \bm{u}_i, \bm{e}_{j^{\prime}}\rangle \in \mathbb{E}}{\exp{(\langle\boldsymbol{u}_i,\bm{e}_{j^{\prime}}\rangle)}}}},
\end{equation}
where $\langle \bm{u}_i,\bm{e}_j\rangle$ is the positive sample and $\langle \bm{u}_i,\bm{e}_{j^{\prime}}\rangle$ is all the samples. $\langle\boldsymbol{u}_i,\bm{e}_j\rangle$ indicates the prediction score of the $(\bm{u}_i,\bm{e}_j)$ pair (\ie dot product in SASRec), where $\boldsymbol{u}_i$ is the user representation as in SASRec.

For the knowledge distillation from teachers, we obtain score distributions $\hat{\text{s}}_{\langle i,j\rangle}$ based on the \emph{instance-level scoring strategy} for in-batch items.
This allows us to compile the score set $\{\hat{\text{s}}_{\langle i,j\rangle}\}$ of in-batch items corresponding to user $i$, where $j\in[1,T]$.
Then, we get the probability distribution of item $j$ corresponding to user $i$ as the \emph{supervised signals} as follows:

\begin{equation}
\text{q}(\bm{e}_j)=\frac{\hat{\text{s}}_{\langle i,j\rangle}}{\sum^T_{k=1}{\hat{s}_{\langle i,k\rangle}}}.
\end{equation}

We can also get an output distribution over target items \wrt user $i$ as follows:

\begin{equation}
\text{p}(\bm{e}_j)=\frac{\exp{(\langle\boldsymbol{u}_i,\bm{e}_j\rangle)}}{\sum_{\langle \bm{u}_i, \bm{e}_{j^{\prime}}\rangle \in \mathbb{E}}{\exp{(\langle\boldsymbol{u}_i,\bm{e}_{j^{\prime}}\rangle)}}}.
\end{equation}

Then, we leverage the \emph{KL-divergence loss} to compare the predicted and supervised probabilities as follows:
\begin{equation}
    \mathcal{L}_{KD}= \sum_{\langle \bm{u}_i,\bm{e}_j\rangle \in \mathbb{P}}{\rm p}(\bm{e}_j)\log{\frac{{\rm p}(\bm{e}_j)}{\text{q}(\bm{e}_j)}},
\end{equation}
where $\langle \bm{u}_i,\bm{e}_j\rangle \in \mathbb{P}$ denote all the $\langle \bm{u}_i,\bm{e}_j\rangle$ pairs that occured in the in-batch positive or negative samples.

Finally, we jointly optimize $\mathcal{L}_{CE}$ and $\mathcal{L}_{KD}$ as follows:
\begin{equation}
    \mathcal{L}=\mathcal{L}_{CE}+\lambda\mathcal{L}_{KD},
\end{equation}
where $\lambda$ is the distillation weight.

\subsubsection{Discussion}
\label{sec:discussion}
We highlight the necessity and advantages of \ourmodel~in practically using PRMs:
\begin{itemize}
    \item Compared with pre-trained recommendation methods, \ourmodel{} can be seamlessly deployed online without any modifications (\ie the online serving pipeline, characteristics and costs are exactly the same as the base recommendation models), achieving a stable, effective and efficient model upgrade. On the contrary, the replacement with PRMs requires heavy engineering efforts and may encounter a period of performance fluctuation and even unexpected issues.
    \item Compared with conventional recommendation methods, \ourmodel{} is able to learn from the rich knowledge encoded in PRMs enchanted by PRMs without any additional online computational overhead, which has been verified especially for cold-start recommendation scenarios.
    \item Compared with existing KD methods in recommendation, our proposed method focuses on integrating knowledge from heterogeneous PRMs across multiple domains. Our KD framework not only integrates the knowledge from multiple PRMs across various domains but also enhances the transferability between domains. By synthesizing information from multiple PRMs, our method captures more complex patterns and dependencies, leading to improved recommendation performance. Furthermore, the integration of multiple PRMs strengthens the generalization and robustness of the knowledge distillation process, effectively addressing the limitations of single-domain KD methods in recognizing and modeling cross-domain interactions.
    \item The philosophy of \ourmodel{} exactly matches those efforts hiring strong PLMs as critics or annotators to enhance smaller models while minimizing changes to the model architecture. In fact, the distilled user-item pseudo interactions serve as a form of synthetic data, which has been verified in various NLP tasks. In summary, \ourmodel{} effectively leverages the knowledge encoded in various PRMs without any additional online deployment cost, following the trend of making full use of high-quality data, and providing an efficient way to practically use PRMs.
\end{itemize}

\section{Experiments}
In this section, we evaluate the effectiveness and efficiency of our proposed method and aim to address the following research questions:
\begin{itemize}
    \item \textbf{RQ1:} How does \ourmodel~perform compared with other pre-trained and conventional recommendation baseline methods?
    \item \textbf{RQ2:} How do the different components of \ourmodel~impact model performance?
    \item \textbf{RQ3:} How does \ourmodel~perform compared to baseline methods in terms of efficiency and universality?
    \item \textbf{RQ4:} How does the hyperparameters, which are  crucial to be considered during the knowledge distillation process, affect \ourmodel?
\end{itemize}

\subsection{Experimental Setup}

\subsubsection{Datasets}
We conduct experiments on five real-world datasets from Amazon as the test sets for evaluation. The statistical details are shown in Table~\blue{\ref{tab:dataset}}.
To simulate real-world scenarios, we directly leverage the same pre-training settings as their original papers for all PRMs (\ie {UniSRec}~\cite{UniSRec2022}, {Recformer}~\cite{Recformer2023}, and {UniM$^2$Rec}~\cite{UniM2Rec2023}) as the teacher models, which are pre-trained via different pre-training datasets. Note that our selected test datasets do not overlap with the pre-training datasets from the above PRMs, which makes the test setting fair comparing to the PRMs and conventional models.

\begin{table}[htbp!] %
	\caption{Statistics of pre-processed target domain datasets for evaluation.}
	\label{tab:dataset}
	\begin{tabular*}{\textwidth}{@{\extracolsep\fill}lrrrrr}
		\toprule
		\textbf{Datasets} & \textbf{\#Users} & \textbf{\#Items} & \textbf{\#Inters.} & \textbf{Avg. $n$} & \textbf{Sparsity} \\
		\midrule

		Instruments  & 24,962 & 9,749 & 207,926 & 8.37 & 99.93\%  \\
  
		Toys &  18,436 & 10,986 & 163,837 & 8.34 & 99.94\%  \\ 

            Scientific & 11,041 & 5,327 & 76,896 & 6.96 & 99.99\% \\

	    Arts & 45,486 & 21,019 & 395,150 & 8.69 & 99.96\% \\

	    Games & 10,235 & 14,798 & 98,374 & 9.03 & 99.89\% \\

		\bottomrule
	\end{tabular*}

    \footnotetext[1]{``Avg. $n$'' denotes the average sequence length.}
\end{table}

\subsubsection{Evaluation Metrics}
For each test set in its corresponding target domain, all PRMs are first tuned on this test set, and then our \ourmodel{} distills knowledge from these PRMs into a student model.
Our proposed approach is evaluated on the next item prediction task and we apply the typical \textsl{leave-one-out} strategy for evaluation as in previous works~\cite{SASRec,UniSRec2022}.
we utilize the typical Recall and NDCG on top-$K$ ranked items to evaluate the performance.
As suggested in~\cite{Metrics2020}, we rank all items for each user in test sets instead of sampling evaluation.

\subsubsection{Baseline Methods}

To demonstrate the effectiveness and efficiency of \ourmodel{}, we compare three kind of baseline methods, \ie pre-trained recommendation models, conventional sequential recommendation models and typical KD in recommendation:

The first kind of methods are the pre-trained recommendation models, which are used as the teacher models in \ourmodel:
\begin{itemize}
    \item \textbf{UniSRec}~\cite{UniSRec2022} employs a PLM with domain adaptors to construct universal item representations from texts and leverages conventional sequential models (\eg SASRec \cite{SASRec}) for sequential behavior modeling.
    \item \textbf{Recformer}~\cite{Recformer2023} treats items as plain texts and employs a behavior-tuned PLM (\ie LongFormer \cite{beltagy2020longformer}) to encode the sequence of historical behaviors, predicting the next item accordingly.
    \item \textbf{UniM$^2$Rec}~\cite{UniM2Rec2023} is designed for recommendation systems, focusing on integrating multiple types of information to improve the accuracy and relevance of recommendations.
\end{itemize}

The second kind of methods are the conventional SR methods:
\begin{itemize}
    \item \textbf{GRU4Rec} \cite{GRU4Rec} adopts RNNs to model user behavior sequences for sequential recommendations.
    \item \textbf{SASRec}~\cite{SASRec} adopts a self-attention network to capture the user’s preference within a sequence.
    \item \textbf{BERT4Rec}~\cite{BERT4Rec} adapts the original text-based BERT model with the cloze objective for modeling user behavior sequences.
    \item \textbf{RecGURU}~\cite{RecGURU} proposes to augment the sequence representations with an auto-encoder via adversarial learning.
\end{itemize}

The third kind of methods are the recommendation knowledge distillation methods:
\begin{itemize}
    \item \textbf{HetComp}~\cite{HetComp2023} proposes leveraging dynamic knowledge construction to provide progressive knowledge transfer.
\end{itemize}

\subsubsection{Implementation Details}

Our proposed \ourmodel{} is implemented in PyTorch with NVIDIA A800 80G.
With regard to fair comparisons,
we set item embedding of \ourmodel{} and competitors with the dimension $D$ of 300,
the attention heads with the number $H$ of 2,
and the layers with the number $L$ of 2 except Recformer, which is based on Longformer.
All trainable parameters in these models are optimized by leveraging Adam with the batch size of 512,
learning rate of 0.001, drop rate of 0.1, and L2 regularisation strength $\lambda$ of $1e-5$. The parameters are the same when comparing them. In the main results, we implement \ourmodel{} based on the representative and widely-used SR baseline named SASRec.

\begin{sidewaystable}
\caption{Main results on multi-domain sequential recommendation.}\label{tab3}
\label{tab:performance}

\renewcommand{\arraystretch}{1.5}
\begin{tabular*}{\textheight}{@{\extracolsep\fill}llcccccccccc}
\toprule%

Dataset & \makebox[0.05\textwidth][c]{Metric} & \makebox[0.05\textwidth][c]{UniSRec} & \makebox[0.05\textwidth][c]{Recformer} & \makebox[0.06\textwidth][c]{UniM$^2$Rec} & \makebox[0.06\textwidth][c]{GRU4Rec} & \makebox[0.06\textwidth][c]{SASRec} & \makebox[0.06\textwidth][c]{BERT4Rec} & \makebox[0.06\textwidth][c]{RecGURU} & \makebox[0.06\textwidth][c]{HetComp} & \textbf{\ourmodel} & \textbf{Improv.} \\
\midrule

Instruments & R@5 & 0.0840 & 0.0835 & 0.0978 & 0.0677 & 0.0860 & 0.0892 & 0.0865 & 0.0879 & \textbf{0.0954}* & $+$10.93\% \\
    
    & R@10 & 0.1104 & 0.1061 & 0.1289 & 0.0843 & 0.0979 & 0.1098 & 0.1082 & 0.0986 &\textbf{0.1089}* & $+$11.24\% \\
    
    & R@20 & 0.1473 & 0.1395 & 0.1697 & 0.1093 & 0.1483 & 0.1572 & 0.1374 & 0.1497 &\textbf{0.1593}* & $+$7.42\% \\

    & N@5 & 0.0548 & 0.0629 & 0.0635 & 0.0496 & 0.0526 & 0.0553 & 0.0594 & 0.0565 &\textbf{0.0597}* & $+$13.50\% \\
 
    & N@10 & 0.0623 & 0.0825 & 0.0759 & 0.0549 & 0.0635 & 0.0631 & 0.0699 & 0.0674 &\textbf{0.0757}* & $+$19.21\% \\

    & N@20 & 0.0717 & 0.0917 & 0.0842 & 0.0612 & 0.0703 & 0.0719 & 0.0811 & 0.0757 &\textbf{0.0814}* & $+$15.79\% \\

Toys & R@5 & 0.1007 & 0.1123 & 0.1053 & 0.0757 & 0.0795 & 0.0694 & 0.0742 & 0.0821 &\textbf{0.0884}* & $+$11.19\% \\
    
    & R@10 & 0.1260 & 0.1341 & 0.1247 & 0.0903 & 0.0979 & 0.0829 & 0.0936 & 0.0995 &\textbf{0.1163}* & $+$18.79\%  \\

    & R@20 & 0.1539 & 0.1574 & 0.1601 & 0.1082 & 0.1184 & 0.0988 & 0.1121 & 0.1209 &\textbf{0.1291}* & $+$9.04\% \\

    & N@5 & 0.0664 & 0.0712 & 0.0692 & 0.0617 & 0.0632 & 0.0560 & 0.0656 & 0.0647 &\textbf{0.0703}* & $+$11.23\%  \\

    & N@10 & 0.0745 & 0.0943 & 0.0901 & 0.0664 & 0.0693 & 0.0604 & 0.0679 & 0.0727 &\textbf{0.0784}* & $+$13.13\% \\

    & N@20 & 0.0816 & 0.0993 & 0.0989 & 0.0709 & 0.0746 & 0.0644 & 0.0763 & 0.0771 &\textbf{0.0819}* & $+$9.79\% \\

Scientific & R@5 & 0.0884 & 0.0875 & 0.0907 & 0.0373 & 0.0735 & 0.0614 & 0.0697 & 0.0723 & \textbf{0.0754}\quad & $+$2.59\% \\

    & R@10 & 0.1215 & 0.1391 & 0.1249 & 0.0588 & 0.1064 & 0.088 & 0.0894 & 0.0989 & \textbf{0.1231}* & $+$15.70\% \\

    & R@20 & 0.1674 & 0.1695 & 0.1723 & 0.0843 & 0.1408 & 0.1221 & 0.1157 & 0.1428 & \textbf{0.1534}* & $+$8.95\% \\

\botrule
\end{tabular*}

\footnotetext[1]{``Improv.'' indicates the relative improvement ratios of \ourmodel~over its base model SASRec.}
\footnotetext[2]{``*'' indicates that the improvements are significant over its corresponding base model ($t$-test with $p<0.05$).}
\end{sidewaystable}

\begin{sidewaystable}
\caption{Continued.}\label{tab3}
\label{tab:performance2}

\renewcommand{\arraystretch}{1.5}
\begin{tabular*}{\textheight}{@{\extracolsep\fill}llcccccccccc}
\toprule%

Dataset & \makebox[0.05\textwidth][c]{Metric} & \makebox[0.05\textwidth][c]{UniSRec} & \makebox[0.05\textwidth][c]{Recformer} & \makebox[0.06\textwidth][c]{UniM$^2$Rec} & \makebox[0.06\textwidth][c]{GRU4Rec} & \makebox[0.06\textwidth][c]{SASRec} & \makebox[0.06\textwidth][c]{BERT4Rec} & \makebox[0.06\textwidth][c]{RecGURU} & \makebox[0.06\textwidth][c]{HetComp} & \textbf{\ourmodel} & \textbf{Improv.} \\
\midrule

    Scientific & N@5 & 0.0553 & 0.0589 & 0.0594 & 0.0242 & 0.0457 & 0.0387 & 0.0513 & 0.0503 &\textbf{0.0534}* & $+$16.85\% \\

    & N@10 & 0.0668 & 0.0764 & 0.0895 & 0.0311 & 0.0549 & 0.0443 & 0.0581 & 0.0570 &\textbf{0.0627}* & $+$14.21\% \\

    & N@20 & 0.0781 & 0.0979 & 0.1012 & 0.0376 & 0.0635 & 0.0502 & 0.0642 & 0.0614 &\textbf{0.0715}* & $+$12.60\% \\

    Arts & R@5 & 0.0905 & 0.1053 & 0.0917 & 0.0565 & 0.0804 & 0.0782 & 0.0797 & 0.0793 & \textbf{0.0879}* & $+$9.33\% \\
    
    & R@10 & 0.1239 & 0.1474 & 0.1304 & 0.0763 & 0.1108 & 0.1039 & 0.1021 & 0.1089 & \textbf{0.1272}* & $+$14.80\% \\

    & R@20 & 0.1665 & 0.1702 & 0.1673 & 0.1028 & 0.1403 & 0.1354 & 0.1320 & 0.1407 & \textbf{0.1490}* &  $+$6.20\% \\

    & N@5 & 0.0570 & 0.0658 & 0.0639 & 0.0405 & 0.0484 & 0.0465 & 0.0621 & 0.0474 &\textbf{0.0495}* & $+$2.27\% \\

    & N@10 & 0.0682 & 0.0797 & 0.0729 & 0.0469 & 0.0587 & 0.0541 & 0.0693 & 0.0592 &\textbf{0.0695}* & $+$18.40\% \\
    
    & N@20 & 0.0786 & 0.0963 & 0.0802 & 0.0535 & 0.0654 & 0.0625 & 0.0768 & 0.0685 &\textbf{0.0729}* & $+$11.47\% \\

    Games & R@5 & 0.0654 & 0.0631 & 0.0679 & 0.0519 & 0.0588 & 0.0567 & 0.0532 & 0.0573 & \textbf{0.0595}\quad & $+$1.19\% \\

    &  R@10 & 0.1021 & 0.1046 & 0.1084 & 0.0844 & 0.0993 & 0.0974 & 0.0951 & 0.0980 &\textbf{0.1097}* & $+$10.47\% \\

    & R@20 & 0.1670 & 0.1598 & 0.1705 & 0.1324 & 0.1516 & 0.1504 & 0.1403 & 0.1492 &\textbf{0.1529}\quad & $+$0.86\%\\

    & N@5 & 0.0366 & 0.0384 & 0.0372 & 0.0335 & 0.0317 & 0.0321 & 0.0341 & 0.0323 &\textbf{0.0348}* & $+$9.78\% \\

    & N@10 & 0.0504 & 0.0569 & 0.0527 & 0.0439 & 0.0445 & 0.0439 & 0.0462 & 0.0451 &\textbf{0.0485}* & $+$8.99\% \\

    & N@20 & 0.0651 & 0.0695 & 0.0673 & 0.0560 & 0.0578 & 0.0557 & 0.0591 & 0.0569 &\textbf{0.0614}* & $+$6.23\% \\

\botrule
\end{tabular*}

\end{sidewaystable}

\subsection{Overall Performance (RQ1)}

We compare our \ourmodel{} with competitive baselines on five target domain test datasets. The results are shown in Table~\ref{tab:performance} and Table~\ref{tab:performance2}. We have following observations:
\begin{itemize}
    \item First, SASRec serves as a strong baseline, consistently outperforming other conventional models across most datasets. This demonstrates the feasibility of selecting SASRec as the student model.
    \item Second, all PRMs outperform conventional SR models on most datasets. This improvement is expected, as PRMs leverage additional pre-training data from other domains and external knowledge derived from PLMs. However, these PRMs incur higher computational and memory costs for PRMs. A comparison of various PRMs reveals that no single PRM consistently outperforms the others across all datasets, particularly when considering differences in PLM sizes and pre-training datasets. Each PRM demonstrates its own strengths and weaknesses depending on the dataset.
    \item Third, we observe that \ourmodel{} provides an effective and efficient solution for fully leveraging different PRMs in multi-domain sequential recommendation.
    \ourmodel{} achieves the best overall performance and demonstrates its superiority over conventional SR methods, while maintaining the same online serving cost as the corresponding base model. The relative improvements in Recall and NDCG, compared to SASRec, range from $0.86\%$ to $19.21\%$, with \ourmodel{} occasionally outperforming all PRMs, further highlighting its effectiveness. 
    \item Finally, the substantial improvements primarily stem from the additional knowledge contributed by the heterogeneous PRMs. On one hand, these PRMs such as UniSRec, Recformer and UniM$^2$Rec predict user preferences for relatively popular items (\eg in-batch negative samples) from different aspects, which helps mitigate the issue of data sparsity in target domains. On the other hand, our proposed method enables an effective and efficient curriculum-scheduled multi-teacher knowledge distillation.
\end{itemize}

\subsection{Ablation Study (RQ2)}
\label{sec.ablation}

In this part, we aim to evaluate how each of the proposed technique affects the final performance.
We design five variants of the proposed \ourmodel~by removing or replacing certain components for comparisons, including:
(\textbf{i}) \underline{$w/o$ MT$_1$}: only using UniSRec as the teacher model.
(\textbf{ii}) \underline{$w/o$ MT$_2$}: only using Recformer as the teacher model.
(\textbf{iii}) \underline{$w/o$ MT$_3$}: only using UniM$^2$Rec as the teacher model.
(\textbf{iv}) \underline{$w/o$ CSS}: without the curriculum-scheduled user behavior sequence sampling strategy.
(\textbf{v}) \underline{$w/o$ IN}: replacing in-batch negative samples with all negative samples in KD.
(\textbf{vi}) \underline{$w/o$ WA}: replacing the instance-level scoring strategy with the fixed weights in the knowledge integration.

\begin{figure}[!htbp]
    \centering
    \includegraphics[width=0.8\linewidth]{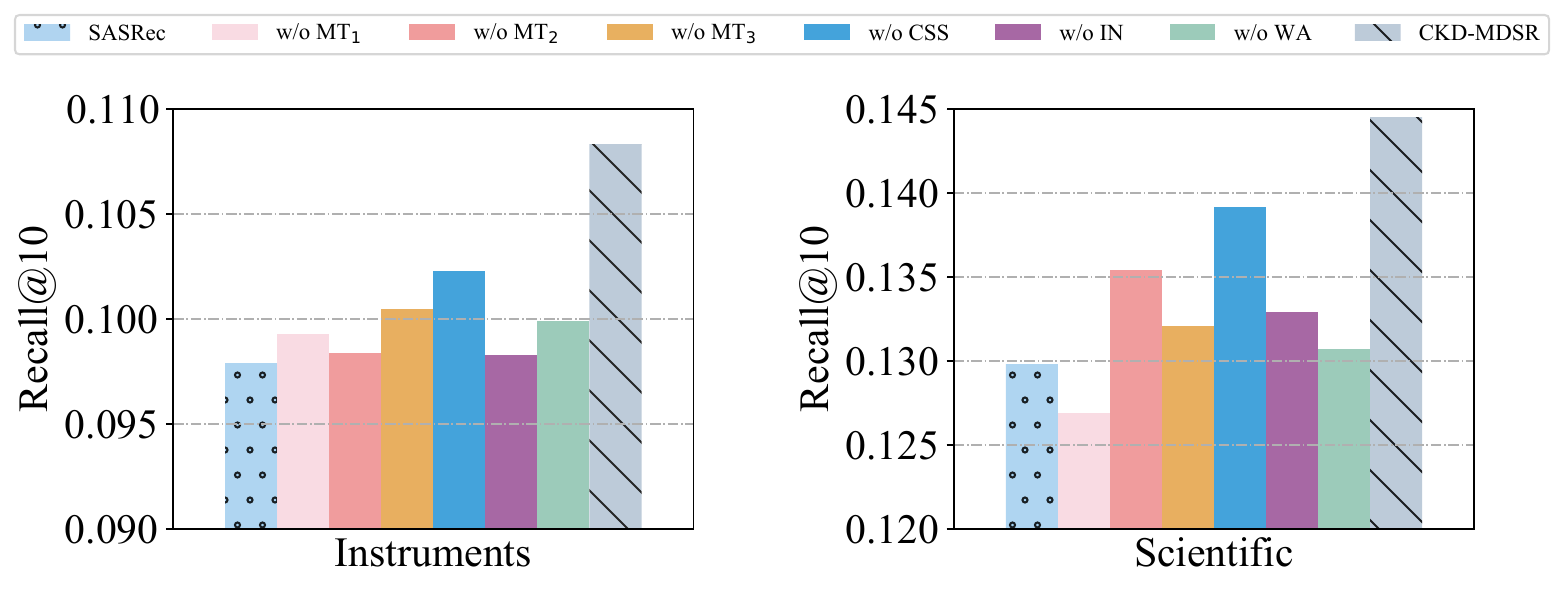}
    \caption{Ablation study of \ourmodel~ on ``Instruments'' and ``Scientific''.}
    \label{fig:ablation}
\end{figure}

The experimental results of the proposed \ourmodel~and its variants are shown in Figure~\blue{\ref{fig:ablation}}, from which we have:
(\textbf{i}) All PRMs are essential in \ourmodel{}.
It indicates that different PRMs can provide diverse but informative knowledge on different user-item pairs.
Besides, it further demonstrates that adopting multi-teacher KD for practically using PRMs is reasonable.
(\textbf{ii}) We find that only considering in-batch negative samples in distillation performs better than distilling PRM scores on all items. The in-batch strategy not only accelerates the distillation process in training, but also improves the model performance (which may imply that even PRMs cannot handle all items).
(\textbf{iii}) The instance-level scoring strategy is critical and necessary in \ourmodel. Due to the performance gaps of PRMs on different datasets, the instance-level scoring strategy could enable a relatively stable multi-teacher distillation to enhance a student. We also find similar results on other datasets.

\subsection{Universality Analysis of \ourmodel~(RQ3)}

In this part, we further conduct experiments \wrt distillation from different teachers to various types of student models to verify the universality of the proposed \ourmodel{}. We have the following observations:
(\textbf{i}) \underline{\ourmodel$_{FM}$}: FM~\cite{FM2010} is a typical context-aware recommendation model based on factorization machines. \ourmodel$_{FM}$ is a variant using \ourmodel{} to distill knowledge from multi-teacher to FM.
(\textbf{ii}) \underline{\ourmodel$_{DF}$}: DeepFM~\cite{DeepFM2017} combines the power of FM and deep learning for more comprehensive feature interaction learning, which is widely used in practice. \ourmodel{}$_{DF}$ uses DeepFM as its corresponding student model. 
(\textbf{iii}) \underline{\ourmodel$_{LG}$}: LightGCN~\cite{LightGCN2020} is a representative graph-based recommendation model verified in different scenarios. \ourmodel$_{LG}$ leverages LightGCN as its corresponding student model.

\begin{figure}[!htbp]
    \centering
    \includegraphics[width=0.8\linewidth]{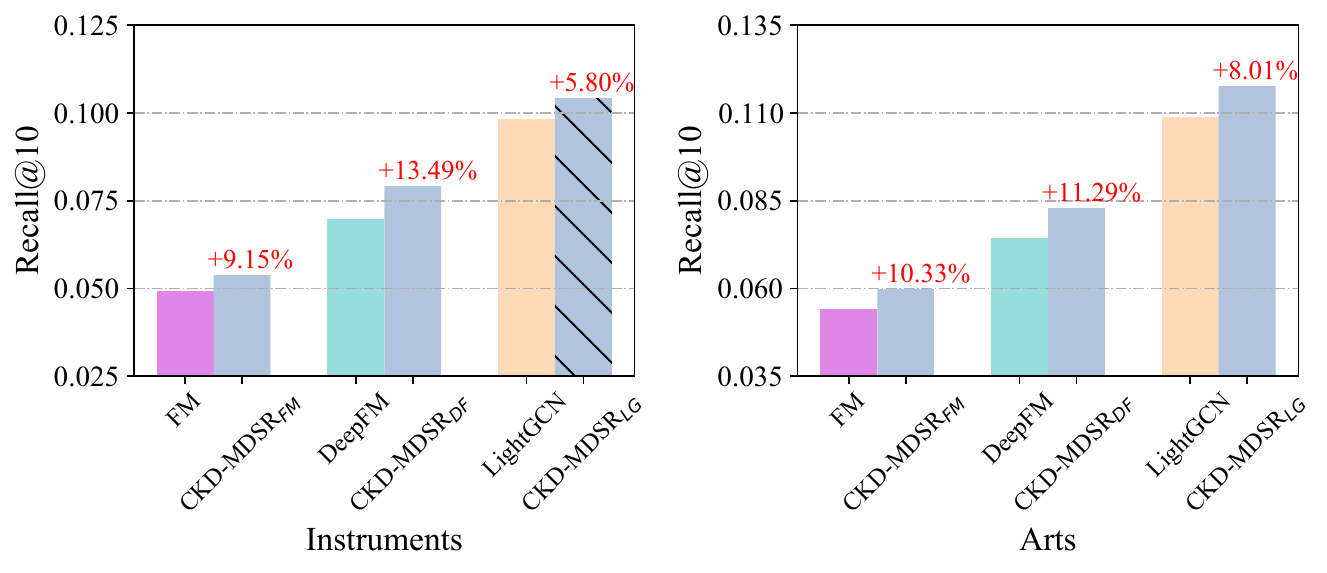}
    \caption{Universality analysis of \ourmodel{} on ``Instruments'' and ``Arts''.}
    \label{fig:universality}
\end{figure}

Fig.~\ref{fig:universality} presents the experimental results regarding the universality of \ourmodel{}.
We have the following observations:
(\textbf{i}) The performance of {\ourmodel$_{FM}$}, {\ourmodel$_{DF}$} and {\ourmodel$_{LG}$} 
are significantly better than {FM}, {DeepFM} and {LightGCN}, respectively on all datasets, indicating that our proposed \ourmodel~is universal and effective to distill knowledge from PRMs to various types of student recommendation models.
(\textbf{ii}) Practical recommender systems comprise various types of models that integrate multi-source information, such as graphs and user/item features, as inputs.
However, existing PRMs are typically built on specific pre-trained model structures, limiting their model generalization to various different downstream models in practical applications.
Additionally, existing PRMs face challenges in directly and reliably replacing online recommendation models due to inherent differences. In contrast, \ourmodel{} facilitates the rapid and effective utilization of pre-trained recommendation models.

\subsection{Analysis on Model Efficiency (RQ3)}

We conduct the empirical study on the model efficiency (\ie inference time cost and memory cost) across different teacher and student models.
To ensure fair comparisons, we maintain consistent experimental settings across all baseline methods (except for Recformer) and report the average results from multiple experimental runs.
Our findings illustrate the trade-off between performance, inference time cost, and memory cost on ``Scientific'' and ``Games'', as shown in Fig.~\ref{fig:efficiency}.
In addition, Table~\ref{tab:efficiency} provides a comprehensive overview of the inference speed and memory cost on ``Arts''.

\begin{table}[!ht]
    \renewcommand\arraystretch{1}
    \setlength{\tabcolsep}{3pt}
    \small
    \centering
    \caption{The results \wrt model efficiency. The metrics are for online inference/serving.}
    \label{tab:efficiency}

    \begin{tabular*}{\textwidth}{@{\extracolsep\fill}lcccc}
        \toprule 
        & \multicolumn{2}{@{}c@{}}{Scientific} & \multicolumn{2}{@{}c@{}}{Arts} \\ \cmidrule{2-3}\cmidrule{4-5}%
        Model & Memory & Time Cost & Memory & Time Cost  \\
        \midrule
        SASRec (Base) &1.0x & 1.0x & 1.0x & 1.0x \\
        UniSRec & 48.3x & 1.8x & 51.5x & 1.9x \\
        Recformer & 78.0x & 2.9x & 81.7x & 3.1x \\
        UniM$^2$Rec & 59.8x & 2.0x & 64.3x & 2.1x \\

        \ourmodel~(Ours) & 1.0x & 1.0x & 1.0x & 1.0x \\

        \bottomrule
    \end{tabular*}

\footnotetext[1]{SASRec is the base model corresponding to ``1.0x'', where ``$n$x'' denotes ``$n$ times''.}
\footnotetext[2]{``Memory'' and ``Time Cost'' denote inference memory and inference time cost, respectively.}

\end{table}

We have the following observations:
(\textbf{i}) \ourmodel{} strikes an optimal balance between performance, inference speed, and memory efficiency, providing an effective means of leveraging PRMs in real-world systems.
(\textbf{ii}) The inference speed and memory cost of \ourmodel{} are comparable to those of SASRec in most cases.
Hence, \ourmodel{} achieves significant performance improvements without incurring additional computational costs.

\begin{figure}[!h]
    \centering
    \includegraphics[width=0.85\linewidth]{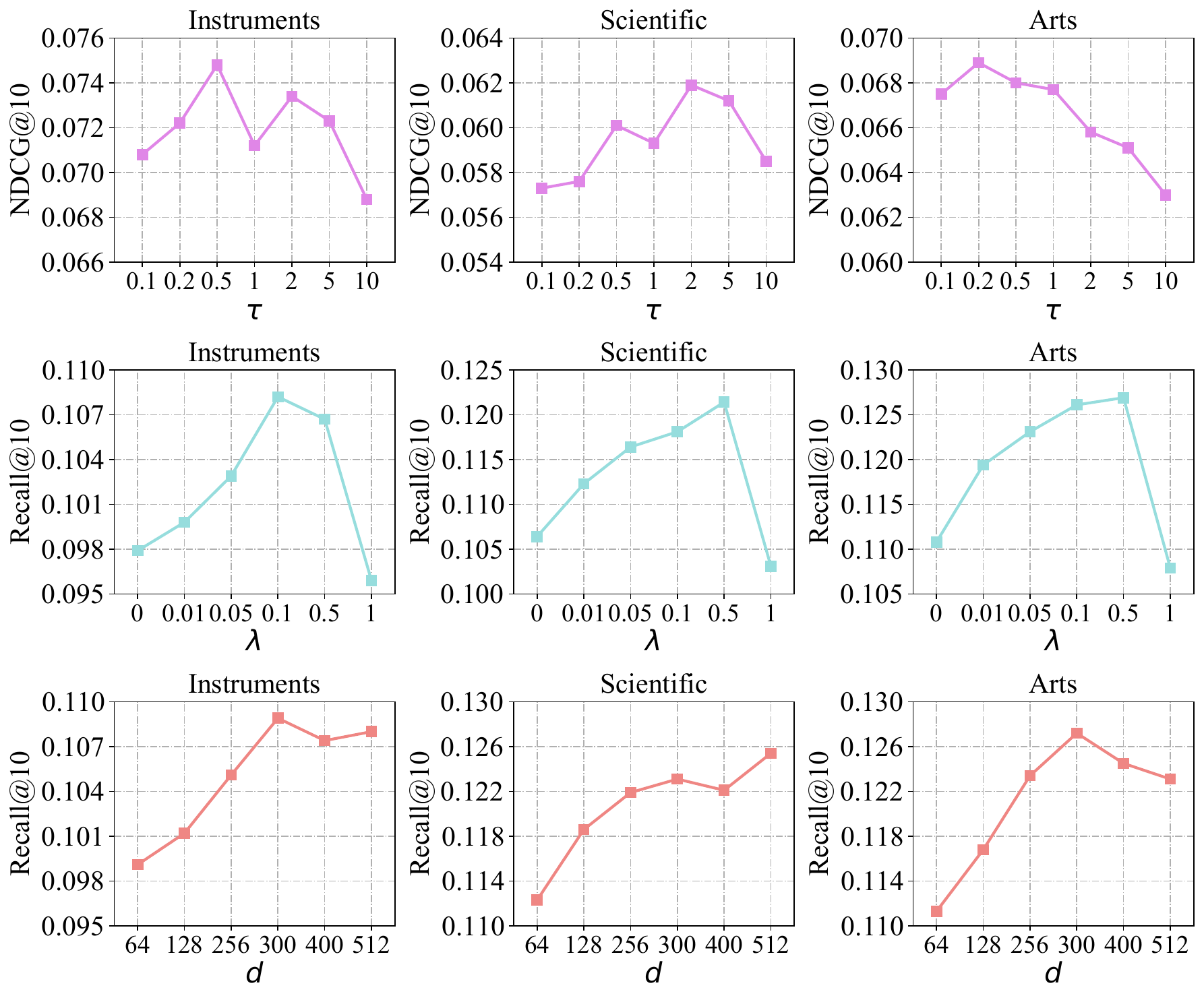}
    \caption{Parameter analyses of \ourmodel~on ``Instruments'', ``Scientific'' and ``Arts''.}
    \label{fig:parameter}
\end{figure}

\subsection{Hyperparameter Sensitivity Analyses (RQ4)}
The results of \ourmodel{} across various hyper-parameters are illustrated in Fig. \ref{fig:parameter}.

\subsubsection{Temperature $\tau$}
We analyze the performance (NDCG@10) across the ``Instruments'', ``Scientific'' and ``Arts'' datasets by varying the temperature parameter $\tau$.
The temperature $\tau$ regulates the level of uncertainty or hardness in the probability distributions during the knowledge distillation process.
It is observed that \ourmodel{} performs optimally when $\tau$ is within the range of $[0.02,2]$ on the ``Instruments'' ``Scientific'' and ``Arts'' datasets.
A smaller $\tau$ results in harder probability distributions, making the model less tolerant of semantically similar samples.
In contrast, a larger $\tau$ yields softer distributions, rendering the model less sensitive to semantic distinctions.

\subsubsection{Distillation Weight $\lambda$}
We analyze the performance (Recall@10) while varying $\lambda$ across the ``Instruments'', ``Scientific'' and ``Arts'' datasets.
The parameter $\lambda$ controls the strength of knowledge distillation.
The experimental results indicate that selecting an appropriate $\lambda$ significantly enhances performance. However, choosing excessively large or small $\lambda$ values can degrade performance. Additionally, the optimal range for $\lambda$ may vary slightly across different datasets.

\subsubsection{Hidden Dimensionality $d$}
We analyze the performance (Recall@10) by varying $d$ across the ``Instruments'', ``Scientific'' and ``Arts'' datasets.
We observe a steady increase in performance as the hidden dimensionality grows, with the best performance achieved at $d=300$.
However, the performance slightly declines when $d=400$, likely due to overfitting.
Thus, $d=300$ often strikes a good balance between efficiency and effectiveness in most datasets.
Additionally, increasing the hidden dimensionality leads to a significant rise in memory overhead.

\section{Conclusion and Future Work}
In this paper, we propose a novel we propose a novel curriculum-scheduled knowledge distillation from multiple pre-trained teachers for multi-domain sequential recommendation to efficiently leverage different pre-trained recommendation models.
In particular, we propose an curriculum-scheduled user behavior sequence sampling strategy to facilitate the model training and an instance-level scoring strategy to select informative knowledge from heterogeneous PRMs.
Different from existing recommendation KD methods that mainly focus on the model simplification, our proposed \ourmodel{} concentrates on integrating knowledge from pre-trained recommendation models into a student model.
With the knowledge distilled from multiple pre-trained recommendation models, we enhance the recommendation performance of the student model without extra inference time and memory cost.
Extensive experiments conducted on five real-world datasets demonstrate the effectiveness, efficiency and universality of our proposed \ourmodel{}.
In real-world systems, our proposed \ourmodel{} is also of critical importance for the practical application of pre-trained recommendation models.
For instance, \ourmodel{} can be utilized to provide online multi-domain sequential recommendation services.

As future work, we aim to explore the dataset distillation to extract the knowledge across multiple modalities and domains from different recommendation datasets.
And we will also rethink the knowledge distillation and ensemble modeling from the perspective of data-centric recommendations.







\bibliographystyle{sn-basic}
\bibliography{sn-bibliography}


\begin{thebibliography}{47}
\ifx \bisbn   \undefined \def \bisbn  #1{ISBN #1}\fi
\ifx \binits  \undefined \def \binits#1{#1}\fi
\ifx \bauthor  \undefined \def \bauthor#1{#1}\fi
\ifx \batitle  \undefined \def \batitle#1{#1}\fi
\ifx \bjtitle  \undefined \def \bjtitle#1{#1}\fi
\ifx \bvolume  \undefined \def \bvolume#1{\textbf{#1}}\fi
\ifx \byear  \undefined \def \byear#1{#1}\fi
\ifx \bissue  \undefined \def \bissue#1{#1}\fi
\ifx \bfpage  \undefined \def \bfpage#1{#1}\fi
\ifx \blpage  \undefined \def \blpage #1{#1}\fi
\ifx \burl  \undefined \def \burl#1{\textsf{#1}}\fi
\ifx \doiurl  \undefined \def \doiurl#1{\url{https://doi.org/#1}}\fi
\ifx \betal  \undefined \def \betal{\textit{et al.}}\fi
\ifx \binstitute  \undefined \def \binstitute#1{#1}\fi
\ifx \binstitutionaled  \undefined \def \binstitutionaled#1{#1}\fi
\ifx \bctitle  \undefined \def \bctitle#1{#1}\fi
\ifx \beditor  \undefined \def \beditor#1{#1}\fi
\ifx \bpublisher  \undefined \def \bpublisher#1{#1}\fi
\ifx \bbtitle  \undefined \def \bbtitle#1{#1}\fi
\ifx \bedition  \undefined \def \bedition#1{#1}\fi
\ifx \bseriesno  \undefined \def \bseriesno#1{#1}\fi
\ifx \blocation  \undefined \def \blocation#1{#1}\fi
\ifx \bsertitle  \undefined \def \bsertitle#1{#1}\fi
\ifx \bsnm \undefined \def \bsnm#1{#1}\fi
\ifx \bsuffix \undefined \def \bsuffix#1{#1}\fi
\ifx \bparticle \undefined \def \bparticle#1{#1}\fi
\ifx \barticle \undefined \def \barticle#1{#1}\fi
\bibcommenthead
\ifx \bconfdate \undefined \def \bconfdate #1{#1}\fi
\ifx \botherref \undefined \def \botherref #1{#1}\fi
\ifx \url \undefined \def \url#1{\textsf{#1}}\fi
\ifx \bchapter \undefined \def \bchapter#1{#1}\fi
\ifx \bbook \undefined \def \bbook#1{#1}\fi
\ifx \bcomment \undefined \def \bcomment#1{#1}\fi
\ifx \oauthor \undefined \def \oauthor#1{#1}\fi
\ifx \citeauthoryear \undefined \def \citeauthoryear#1{#1}\fi
\ifx \endbibitem  \undefined \def \endbibitem {}\fi
\ifx \bconflocation  \undefined \def \bconflocation#1{#1}\fi
\ifx \arxivurl  \undefined \def \arxivurl#1{\textsf{#1}}\fi
\csname PreBibitemsHook\endcsname

\bibitem[\protect\citeauthoryear{Hou et~al.}{2022}]{UniSRec2022}
\begin{bchapter}
\bauthor{\bsnm{Hou}, \binits{Y.}},
\bauthor{\bsnm{Mu}, \binits{S.}},
\bauthor{\bsnm{Zhao}, \binits{W.X.}},
\bauthor{\bsnm{Li}, \binits{Y.}},
\bauthor{\bsnm{Ding}, \binits{B.}},
\bauthor{\bsnm{Wen}, \binits{J.}}:
\bctitle{Towards universal sequence representation learning for recommender systems}.
In: \bbtitle{{KDD} '22: The 28th {ACM} {SIGKDD} Conference on Knowledge Discovery and Data Mining, Washington, DC, USA, August 14 - 18, 2022},
pp. \bfpage{585}--\blpage{593}
(\byear{2022})
\end{bchapter}
\endbibitem

\bibitem[\protect\citeauthoryear{Li et~al.}{2023}]{Recformer2023}
\begin{bchapter}
\bauthor{\bsnm{Li}, \binits{J.}},
\bauthor{\bsnm{Wang}, \binits{M.}},
\bauthor{\bsnm{Li}, \binits{J.}},
\bauthor{\bsnm{Fu}, \binits{J.}},
\bauthor{\bsnm{Shen}, \binits{X.}},
\bauthor{\bsnm{Shang}, \binits{J.}},
\bauthor{\bsnm{McAuley}, \binits{J.}}:
\bctitle{Text is all you need: Learning language representations for sequential recommendation}.
In: \bbtitle{KDD}
(\byear{2023})
\end{bchapter}
\endbibitem

\bibitem[\protect\citeauthoryear{Yuan et~al.}{2020}]{PeterRec2020}
\begin{bchapter}
\bauthor{\bsnm{Yuan}, \binits{F.}},
\bauthor{\bsnm{He}, \binits{X.}},
\bauthor{\bsnm{Karatzoglou}, \binits{A.}},
\bauthor{\bsnm{Zhang}, \binits{L.}}:
\bctitle{Parameter-efficient transfer from sequential behaviors for user modeling and recommendation}.
In: \bbtitle{SIGIR}
(\byear{2020})
\end{bchapter}
\endbibitem

\bibitem[\protect\citeauthoryear{Qiu et~al.}{2021}]{qiu2021U-BERT}
\begin{bchapter}
\bauthor{\bsnm{Qiu}, \binits{Z.}},
\bauthor{\bsnm{Wu}, \binits{X.}},
\bauthor{\bsnm{Gao}, \binits{J.}},
\bauthor{\bsnm{Fan}, \binits{W.}}:
\bctitle{U-bert: Pre-training user representations for improved recommendation}.
In: \bbtitle{AAAI}
(\byear{2021})
\end{bchapter}
\endbibitem

\bibitem[\protect\citeauthoryear{Mao et~al.}{2023}]{mao2023unitrec}
\begin{botherref}
\oauthor{\bsnm{Mao}, \binits{Z.}},
\oauthor{\bsnm{Wang}, \binits{H.}},
\oauthor{\bsnm{Du}, \binits{Y.}},
\oauthor{\bsnm{Wong}, \binits{K.-f.}}:
Unitrec: A unified text-to-text transformer and joint contrastive learning framework for text-based recommendation.
arXiv preprint arXiv:2305.15756
(2023)
\end{botherref}
\endbibitem

\bibitem[\protect\citeauthoryear{Wang et~al.}{2023}]{MISSRec2023}
\begin{bchapter}
\bauthor{\bsnm{Wang}, \binits{J.}},
\bauthor{\bsnm{Zeng}, \binits{Z.}},
\bauthor{\bsnm{Wang}, \binits{Y.}},
\bauthor{\bsnm{Wang}, \binits{Y.}},
\bauthor{\bsnm{Lu}, \binits{X.}},
\bauthor{\bsnm{Li}, \binits{T.}},
\bauthor{\bsnm{Yuan}, \binits{J.}},
\bauthor{\bsnm{Zhang}, \binits{R.}},
\bauthor{\bsnm{Zheng}, \binits{H.}},
\bauthor{\bsnm{Xia}, \binits{S.}}:
\bctitle{Missrec: Pre-training and transferring multi-modal interest-aware sequence representation for recommendation}.
In: \bbtitle{Proceedings of the 31st {ACM} International Conference on Multimedia, {MM} 2023, Ottawa, ON, Canada, 29 October 2023- 3 November 2023}
(\byear{2023})
\end{bchapter}
\endbibitem

\bibitem[\protect\citeauthoryear{Kang and McAuley}{2018}]{SASRec}
\begin{bchapter}
\bauthor{\bsnm{Kang}, \binits{W.}},
\bauthor{\bsnm{McAuley}, \binits{J.J.}}:
\bctitle{Self-attentive sequential recommendation}.
In: \bbtitle{ICDM}
(\byear{2018})
\end{bchapter}
\endbibitem

\bibitem[\protect\citeauthoryear{Sun et~al.}{2023}]{UniM2Rec2023}
\begin{botherref}
\oauthor{\bsnm{Sun}, \binits{W.}},
\oauthor{\bsnm{Xie}, \binits{R.}},
\oauthor{\bsnm{Bian}, \binits{S.}},
\oauthor{\bsnm{Zhao}, \binits{W.X.}},
\oauthor{\bsnm{Zhou}, \binits{J.}}:
Universal multi-modal multi-domain pre-trained recommendation.
arXiv preprint arXiv:2311.01831
(2023)
\end{botherref}
\endbibitem

\bibitem[\protect\citeauthoryear{Li et~al.}{2021}]{lightweight2021}
\begin{bchapter}
\bauthor{\bsnm{Li}, \binits{Y.}},
\bauthor{\bsnm{Chen}, \binits{T.}},
\bauthor{\bsnm{Zhang}, \binits{P.}},
\bauthor{\bsnm{Yin}, \binits{H.}}:
\bctitle{Lightweight self-attentive sequential recommendation}.
In: \bbtitle{CIKM}
(\byear{2021})
\end{bchapter}
\endbibitem

\bibitem[\protect\citeauthoryear{Zhou et~al.}{2022}]{FMLP-Rec2022}
\begin{bchapter}
\bauthor{\bsnm{Zhou}, \binits{K.}},
\bauthor{\bsnm{Yu}, \binits{H.}},
\bauthor{\bsnm{Zhao}, \binits{W.X.}},
\bauthor{\bsnm{Wen}, \binits{J.}}:
\bctitle{Filter-enhanced {MLP} is all you need for sequential recommendation}.
In: \bbtitle{WWW}
(\byear{2022})
\end{bchapter}
\endbibitem

\bibitem[\protect\citeauthoryear{Hinton et~al.}{2015}]{hinton2015distilling}
\begin{botherref}
\oauthor{\bsnm{Hinton}, \binits{G.}},
\oauthor{\bsnm{Vinyals}, \binits{O.}},
\oauthor{\bsnm{Dean}, \binits{J.}}:
Distilling the knowledge in a neural network.
arXiv preprint arXiv:1503.02531
(2015)
\end{botherref}
\endbibitem

\bibitem[\protect\citeauthoryear{Chen et~al.}{2023}]{UnbiasedKD2023}
\begin{bchapter}
\bauthor{\bsnm{Chen}, \binits{G.}},
\bauthor{\bsnm{Chen}, \binits{J.}},
\bauthor{\bsnm{Feng}, \binits{F.}},
\bauthor{\bsnm{Zhou}, \binits{S.}},
\bauthor{\bsnm{He}, \binits{X.}}:
\bctitle{Unbiased knowledge distillation for recommendation}.
In: \bbtitle{WSDM}
(\byear{2023})
\end{bchapter}
\endbibitem

\bibitem[\protect\citeauthoryear{Lin et~al.}{2023}]{huaweirec2023}
\begin{botherref}
\oauthor{\bsnm{Lin}, \binits{J.}},
\oauthor{\bsnm{Dai}, \binits{X.}},
\oauthor{\bsnm{Xi}, \binits{Y.}},
\oauthor{\bsnm{Liu}, \binits{W.}},
\oauthor{\bsnm{Chen}, \binits{B.}},
\oauthor{\bsnm{Li}, \binits{X.}},
\oauthor{\bsnm{Zhu}, \binits{C.}},
\oauthor{\bsnm{Guo}, \binits{H.}},
\oauthor{\bsnm{Yu}, \binits{Y.}},
\oauthor{\bsnm{Tang}, \binits{R.}}, et al.:
How can recommender systems benefit from large language models: A survey.
arXiv preprint arXiv:2306.05817
(2023)
\end{botherref}
\endbibitem

\bibitem[\protect\citeauthoryear{Wei et~al.}{2024}]{LLMRec2024}
\begin{bchapter}
\bauthor{\bsnm{Wei}, \binits{W.}},
\bauthor{\bsnm{Ren}, \binits{X.}},
\bauthor{\bsnm{Tang}, \binits{J.}},
\bauthor{\bsnm{Wang}, \binits{Q.}},
\bauthor{\bsnm{Su}, \binits{L.}},
\bauthor{\bsnm{Cheng}, \binits{S.}},
\bauthor{\bsnm{Wang}, \binits{J.}},
\bauthor{\bsnm{Yin}, \binits{D.}},
\bauthor{\bsnm{Huang}, \binits{C.}}:
\bctitle{Llmrec: Large language models with graph augmentation for recommendation}.
In: \bbtitle{WSDM}
(\byear{2024})
\end{bchapter}
\endbibitem

\bibitem[\protect\citeauthoryear{Sun et~al.}{2023}]{WWWJ-sequential-recommendation}
\begin{barticle}
\bauthor{\bsnm{Sun}, \binits{K.}},
\bauthor{\bsnm{Qian}, \binits{T.}},
\bauthor{\bsnm{Zhong}, \binits{M.}},
\bauthor{\bsnm{Li}, \binits{X.}}:
\batitle{Towards more effective encoders in pre-training for sequential recommendation}.
\bjtitle{World Wide Web {(WWW)}}
\bvolume{26}(\bissue{5}),
\bfpage{2801}--\blpage{2832}
(\byear{2023})
\end{barticle}
\endbibitem

\bibitem[\protect\citeauthoryear{Beltagy et~al.}{2020}]{beltagy2020longformer}
\begin{botherref}
\oauthor{\bsnm{Beltagy}, \binits{I.}},
\oauthor{\bsnm{Peters}, \binits{M.E.}},
\oauthor{\bsnm{Cohan}, \binits{A.}}:
Longformer: The long-document transformer.
arXiv preprint arXiv:2004.05150
(2020)
\end{botherref}
\endbibitem

\bibitem[\protect\citeauthoryear{Sheng et~al.}{2021}]{STAR2021}
\begin{bchapter}
\bauthor{\bsnm{Sheng}, \binits{X.}},
\bauthor{\bsnm{Zhao}, \binits{L.}},
\bauthor{\bsnm{Zhou}, \binits{G.}},
\bauthor{\bsnm{Ding}, \binits{X.}},
\bauthor{\bsnm{Dai}, \binits{B.}},
\bauthor{\bsnm{Luo}, \binits{Q.}},
\bauthor{\bsnm{Yang}, \binits{S.}},
\bauthor{\bsnm{Lv}, \binits{J.}},
\bauthor{\bsnm{Zhang}, \binits{C.}},
\bauthor{\bsnm{Deng}, \binits{H.}},
\bauthor{\bsnm{Zhu}, \binits{X.}}:
\bctitle{One model to serve all: Star topology adaptive recommender for multi-domain {CTR} prediction}.
In: \bbtitle{{CIKM} '21: The 30th {ACM} International Conference on Information and Knowledge Management, Virtual Event, Queensland, Australia, November 1 - 5, 2021},
pp. \bfpage{4104}--\blpage{4113}
(\byear{2021})
\end{bchapter}
\endbibitem

\bibitem[\protect\citeauthoryear{Hwang et~al.}{2024}]{MDSR-2024}
\begin{bchapter}
\bauthor{\bsnm{Hwang}, \binits{J.}},
\bauthor{\bsnm{Ju}, \binits{H.}},
\bauthor{\bsnm{Kang}, \binits{S.}},
\bauthor{\bsnm{Jang}, \binits{S.}},
\bauthor{\bsnm{Yu}, \binits{H.}}:
\bctitle{Multi-domain sequential recommendation via domain space learning}.
In: \bbtitle{Proceedings of the 47th International {ACM} {SIGIR} Conference on Research and Development in Information Retrieval, {SIGIR} 2024, Washington DC, USA, July 14-18, 2024},
pp. \bfpage{2134}--\blpage{2144}
(\byear{2024})
\end{bchapter}
\endbibitem

\bibitem[\protect\citeauthoryear{Chen et~al.}{2024}]{WWWJ-Cross-domain}
\begin{barticle}
\bauthor{\bsnm{Chen}, \binits{J.}},
\bauthor{\bsnm{Zhang}, \binits{F.}},
\bauthor{\bsnm{Li}, \binits{H.}},
\bauthor{\bsnm{Lu}, \binits{H.}},
\bauthor{\bsnm{Jin}, \binits{X.}},
\bauthor{\bsnm{Liu}, \binits{K.}},
\bauthor{\bsnm{Li}, \binits{H.}},
\bauthor{\bsnm{Wang}, \binits{Y.}}:
\batitle{Empnet: An extract-map-predict neural network architecture for cross-domain recommendation}.
\bjtitle{World Wide Web {(WWW)}}
\bvolume{27}(\bissue{2}),
\bfpage{12}
(\byear{2024})
\end{barticle}
\endbibitem

\bibitem[\protect\citeauthoryear{Hao et~al.}{2021}]{AFT2021}
\begin{bchapter}
\bauthor{\bsnm{Hao}, \binits{X.}},
\bauthor{\bsnm{Liu}, \binits{Y.}},
\bauthor{\bsnm{Xie}, \binits{R.}},
\bauthor{\bsnm{Ge}, \binits{K.}},
\bauthor{\bsnm{Tang}, \binits{L.}},
\bauthor{\bsnm{Zhang}, \binits{X.}},
\bauthor{\bsnm{Lin}, \binits{L.}}:
\bctitle{Adversarial feature translation for multi-domain recommendation}.
In: \bbtitle{{KDD} '21: The 27th {ACM} {SIGKDD} Conference on Knowledge Discovery and Data Mining, Virtual Event, Singapore, August 14-18, 2021},
pp. \bfpage{2964}--\blpage{2973}
(\byear{2021})
\end{bchapter}
\endbibitem

\bibitem[\protect\citeauthoryear{Jiang et~al.}{2022}]{ADIN2022}
\begin{bchapter}
\bauthor{\bsnm{Jiang}, \binits{Y.}},
\bauthor{\bsnm{Li}, \binits{Q.}},
\bauthor{\bsnm{Zhu}, \binits{H.}},
\bauthor{\bsnm{Yu}, \binits{J.}},
\bauthor{\bsnm{Li}, \binits{J.}},
\bauthor{\bsnm{Xu}, \binits{Z.}},
\bauthor{\bsnm{Dong}, \binits{H.}},
\bauthor{\bsnm{Zheng}, \binits{B.}}:
\bctitle{Adaptive domain interest network for multi-domain recommendation}.
In: \bbtitle{Proceedings of the 31st {ACM} International Conference on Information {\&} Knowledge Management, Atlanta, GA, USA, October 17-21, 2022},
pp. \bfpage{3212}--\blpage{3221}
(\byear{2022})
\end{bchapter}
\endbibitem

\bibitem[\protect\citeauthoryear{Ma et~al.}{2019}]{PINet2019}
\begin{bchapter}
\bauthor{\bsnm{Ma}, \binits{M.}},
\bauthor{\bsnm{Ren}, \binits{P.}},
\bauthor{\bsnm{Lin}, \binits{Y.}},
\bauthor{\bsnm{Chen}, \binits{Z.}},
\bauthor{\bsnm{Ma}, \binits{J.}},
\bauthor{\bsnm{Rijke}, \binits{M.}}:
\bctitle{{\(\pi\)}-net: {A} parallel information-sharing network for shared-account cross-domain sequential recommendations}.
In: \bbtitle{Proceedings of the 42nd International {ACM} {SIGIR} Conference on Research and Development in Information Retrieval, {SIGIR} 2019, Paris, France, July 21-25, 2019},
pp. \bfpage{685}--\blpage{694}
(\byear{2019})
\end{bchapter}
\endbibitem

\bibitem[\protect\citeauthoryear{Sun et~al.}{2023}]{PSJNet2023}
\begin{barticle}
\bauthor{\bsnm{Sun}, \binits{W.}},
\bauthor{\bsnm{Ma}, \binits{M.}},
\bauthor{\bsnm{Ren}, \binits{P.}},
\bauthor{\bsnm{Lin}, \binits{Y.}},
\bauthor{\bsnm{Chen}, \binits{Z.}},
\bauthor{\bsnm{Ren}, \binits{Z.}},
\bauthor{\bsnm{Ma}, \binits{J.}},
\bauthor{\bsnm{Rijke}, \binits{M.}}:
\batitle{Parallel split-join networks for shared account cross-domain sequential recommendations}.
\bjtitle{{IEEE} Trans. Knowl. Data Eng.}
\bvolume{35}(\bissue{4}),
\bfpage{4106}--\blpage{4123}
(\byear{2023})
\end{barticle}
\endbibitem

\bibitem[\protect\citeauthoryear{Xie et~al.}{2022}]{CCDR2022}
\begin{bchapter}
\bauthor{\bsnm{Xie}, \binits{R.}},
\bauthor{\bsnm{Liu}, \binits{Q.}},
\bauthor{\bsnm{Wang}, \binits{L.}},
\bauthor{\bsnm{Liu}, \binits{S.}},
\bauthor{\bsnm{Zhang}, \binits{B.}},
\bauthor{\bsnm{Lin}, \binits{L.}}:
\bctitle{Contrastive cross-domain recommendation in matching}.
In: \bbtitle{{KDD} '22: The 28th {ACM} {SIGKDD} Conference on Knowledge Discovery and Data Mining, Washington, DC, USA, August 14 - 18, 2022},
pp. \bfpage{4226}--\blpage{4236}
(\byear{2022})
\end{bchapter}
\endbibitem

\bibitem[\protect\citeauthoryear{Hsieh et~al.}{2023}]{DistillStep2023}
\begin{bchapter}
\bauthor{\bsnm{Hsieh}, \binits{C.}},
\bauthor{\bsnm{Li}, \binits{C.}},
\bauthor{\bsnm{Yeh}, \binits{C.}},
\bauthor{\bsnm{Nakhost}, \binits{H.}},
\bauthor{\bsnm{Fujii}, \binits{Y.}},
\bauthor{\bsnm{Ratner}, \binits{A.}},
\bauthor{\bsnm{Krishna}, \binits{R.}},
\bauthor{\bsnm{Lee}, \binits{C.}},
\bauthor{\bsnm{Pfister}, \binits{T.}}:
\bctitle{Distilling step-by-step! outperforming larger language models with less training data and smaller model sizes}.
In: \bbtitle{Findings of ACL}
(\byear{2023})
\end{bchapter}
\endbibitem

\bibitem[\protect\citeauthoryear{Lee and Kim}{2021}]{DualDistill2021}
\begin{bchapter}
\bauthor{\bsnm{Lee}, \binits{Y.}},
\bauthor{\bsnm{Kim}, \binits{K.}}:
\bctitle{Dual correction strategy for ranking distillation in top-n recommender system}.
In: \bbtitle{CIKM}
(\byear{2021})
\end{bchapter}
\endbibitem

\bibitem[\protect\citeauthoryear{Kang et~al.}{2022}]{PnKD2022}
\begin{botherref}
\oauthor{\bsnm{Kang}, \binits{S.}},
\oauthor{\bsnm{Lee}, \binits{D.}},
\oauthor{\bsnm{Kweon}, \binits{W.}},
\oauthor{\bsnm{Yu}, \binits{H.}}:
Personalized knowledge distillation for recommender system.
Knowl. Based Syst.
(2022)
\end{botherref}
\endbibitem

\bibitem[\protect\citeauthoryear{Chen et~al.}{2019}]{AdversarialDistillation2019}
\begin{botherref}
\oauthor{\bsnm{Chen}, \binits{X.}},
\oauthor{\bsnm{Zhang}, \binits{Y.}},
\oauthor{\bsnm{Xu}, \binits{H.}},
\oauthor{\bsnm{Qin}, \binits{Z.}},
\oauthor{\bsnm{Zha}, \binits{H.}}:
Adversarial distillation for efficient recommendation with external knowledge.
{ACM} Trans. Inf. Syst.
(2019)
\end{botherref}
\endbibitem

\bibitem[\protect\citeauthoryear{Tang and Wang}{2018}]{RankingDistillation2018}
\begin{bchapter}
\bauthor{\bsnm{Tang}, \binits{J.}},
\bauthor{\bsnm{Wang}, \binits{K.}}:
\bctitle{Ranking distillation: Learning compact ranking models with high performance for recommender system}.
In: \bbtitle{KDD}
(\byear{2018})
\end{bchapter}
\endbibitem

\bibitem[\protect\citeauthoryear{Kweon et~al.}{2021}]{BidirectionalKD2021}
\begin{bchapter}
\bauthor{\bsnm{Kweon}, \binits{W.}},
\bauthor{\bsnm{Kang}, \binits{S.}},
\bauthor{\bsnm{Yu}, \binits{H.}}:
\bctitle{Bidirectional distillation for top-k recommender system}.
In: \bbtitle{WWW}
(\byear{2021})
\end{bchapter}
\endbibitem

\bibitem[\protect\citeauthoryear{Wang et~al.}{2019}]{CFKD2019}
\begin{bchapter}
\bauthor{\bsnm{Wang}, \binits{H.}},
\bauthor{\bsnm{Lian}, \binits{D.}},
\bauthor{\bsnm{Ge}, \binits{Y.}}:
\bctitle{Binarized collaborative filtering with distilling graph convolutional network}.
In: \bbtitle{IJCAI}
(\byear{2019})
\end{bchapter}
\endbibitem

\bibitem[\protect\citeauthoryear{Kang et~al.}{2020}]{KDRS2020}
\begin{bchapter}
\bauthor{\bsnm{Kang}, \binits{S.}},
\bauthor{\bsnm{Hwang}, \binits{J.}},
\bauthor{\bsnm{Kweon}, \binits{W.}},
\bauthor{\bsnm{Yu}, \binits{H.}}:
\bctitle{{DE-RRD:} {A} knowledge distillation framework for recommender system}.
In: \bbtitle{CIKM}
(\byear{2020})
\end{bchapter}
\endbibitem

\bibitem[\protect\citeauthoryear{Kang et~al.}{2023}]{HetComp2023}
\begin{bchapter}
\bauthor{\bsnm{Kang}, \binits{S.}},
\bauthor{\bsnm{Kweon}, \binits{W.}},
\bauthor{\bsnm{Lee}, \binits{D.}},
\bauthor{\bsnm{Lian}, \binits{J.}},
\bauthor{\bsnm{Xie}, \binits{X.}},
\bauthor{\bsnm{Yu}, \binits{H.}}:
\bctitle{Distillation from heterogeneous models for top-k recommendation}.
In: \bbtitle{Proceedings of the {ACM} Web Conference 2023, {WWW} 2023, Austin, TX, USA, 30 April 2023 - 4 May 2023},
pp. \bfpage{801}--\blpage{811}
(\byear{2023})
\end{bchapter}
\endbibitem

\bibitem[\protect\citeauthoryear{Du et~al.}{2023}]{EMKD2023}
\begin{bchapter}
\bauthor{\bsnm{Du}, \binits{H.}},
\bauthor{\bsnm{Yuan}, \binits{H.}},
\bauthor{\bsnm{Zhao}, \binits{P.}},
\bauthor{\bsnm{Zhuang}, \binits{F.}},
\bauthor{\bsnm{Liu}, \binits{G.}},
\bauthor{\bsnm{Zhao}, \binits{L.}},
\bauthor{\bsnm{Liu}, \binits{Y.}},
\bauthor{\bsnm{Sheng}, \binits{V.S.}}:
\bctitle{Ensemble modeling with contrastive knowledge distillation for sequential recommendation}.
In: \bbtitle{SIGIR}
(\byear{2023})
\end{bchapter}
\endbibitem

\bibitem[\protect\citeauthoryear{Geng et~al.}{2022}]{P5-2022}
\begin{bchapter}
\bauthor{\bsnm{Geng}, \binits{S.}},
\bauthor{\bsnm{Liu}, \binits{S.}},
\bauthor{\bsnm{Fu}, \binits{Z.}},
\bauthor{\bsnm{Ge}, \binits{Y.}},
\bauthor{\bsnm{Zhang}, \binits{Y.}}:
\bctitle{Recommendation as language processing {(RLP):} {A} unified pretrain, personalized prompt {\&} predict paradigm {(P5)}}.
In: \bbtitle{RecSys}
(\byear{2022})
\end{bchapter}
\endbibitem

\bibitem[\protect\citeauthoryear{Hou et~al.}{2023}]{hou2023large}
\begin{botherref}
\oauthor{\bsnm{Hou}, \binits{Y.}},
\oauthor{\bsnm{Zhang}, \binits{J.}},
\oauthor{\bsnm{Lin}, \binits{Z.}},
\oauthor{\bsnm{Lu}, \binits{H.}},
\oauthor{\bsnm{Xie}, \binits{R.}},
\oauthor{\bsnm{McAuley}, \binits{J.}},
\oauthor{\bsnm{Zhao}, \binits{W.X.}}:
Large language models are zero-shot rankers for recommender systems.
arXiv preprint arXiv:2305.08845
(2023)
\end{botherref}
\endbibitem

\bibitem[\protect\citeauthoryear{Chen et~al.}{2021}]{CL_rec}
\begin{bchapter}
\bauthor{\bsnm{Chen}, \binits{Y.}},
\bauthor{\bsnm{Wang}, \binits{X.}},
\bauthor{\bsnm{Fan}, \binits{M.}},
\bauthor{\bsnm{Huang}, \binits{J.}},
\bauthor{\bsnm{Yang}, \binits{S.}},
\bauthor{\bsnm{Zhu}, \binits{W.}}:
\bctitle{Curriculum meta-learning for next poi recommendation}.
In: \bbtitle{Proceedings of the 27th ACM SIGKDD Conference on Knowledge Discovery \& Data Mining},
pp. \bfpage{2692}--\blpage{2702}
(\byear{2021})
\end{bchapter}
\endbibitem

\bibitem[\protect\citeauthoryear{Hacohen and Weinshall}{2019}]{CL_DNN}
\begin{bchapter}
\bauthor{\bsnm{Hacohen}, \binits{G.}},
\bauthor{\bsnm{Weinshall}, \binits{D.}}:
\bctitle{On the power of curriculum learning in training deep networks}.
In: \bbtitle{International Conference on Machine Learning},
pp. \bfpage{2535}--\blpage{2544}
(\byear{2019}).
\bcomment{PMLR}
\end{bchapter}
\endbibitem

\bibitem[\protect\citeauthoryear{Xu et~al.}{2020}]{CL_NLP}
\begin{bchapter}
\bauthor{\bsnm{Xu}, \binits{B.}},
\bauthor{\bsnm{Zhang}, \binits{L.}},
\bauthor{\bsnm{Mao}, \binits{Z.}},
\bauthor{\bsnm{Wang}, \binits{Q.}},
\bauthor{\bsnm{Xie}, \binits{H.}},
\bauthor{\bsnm{Zhang}, \binits{Y.}}:
\bctitle{Curriculum learning for natural language understanding}.
In: \bbtitle{Proceedings of the 58th Annual Meeting of the Association for Computational Linguistics},
pp. \bfpage{6095}--\blpage{6104}
(\byear{2020})
\end{bchapter}
\endbibitem

\bibitem[\protect\citeauthoryear{Penha and Hauff}{2019}]{CL_IR}
\begin{botherref}
\oauthor{\bsnm{Penha}, \binits{G.}},
\oauthor{\bsnm{Hauff}, \binits{C.}}:
Curriculum learning strategies for ir: An empirical study on conversation response ranking.
arXiv preprint arXiv:1912.08555
(2019)
\end{botherref}
\endbibitem

\bibitem[\protect\citeauthoryear{Krichene and Rendle}{2020}]{Metrics2020}
\begin{bchapter}
\bauthor{\bsnm{Krichene}, \binits{W.}},
\bauthor{\bsnm{Rendle}, \binits{S.}}:
\bctitle{On sampled metrics for item recommendation}.
In: \bbtitle{KDD}
(\byear{2020})
\end{bchapter}
\endbibitem

\bibitem[\protect\citeauthoryear{Hidasi et~al.}{2016}]{GRU4Rec}
\begin{bchapter}
\bauthor{\bsnm{Hidasi}, \binits{B.}},
\bauthor{\bsnm{Karatzoglou}, \binits{A.}},
\bauthor{\bsnm{Baltrunas}, \binits{L.}},
\bauthor{\bsnm{Tikk}, \binits{D.}}:
\bctitle{Session-based recommendations with recurrent neural networks}.
In: \bbtitle{ICLR}
(\byear{2016})
\end{bchapter}
\endbibitem

\bibitem[\protect\citeauthoryear{Sun et~al.}{2019}]{BERT4Rec}
\begin{bchapter}
\bauthor{\bsnm{Sun}, \binits{F.}},
\bauthor{\bsnm{Liu}, \binits{J.}},
\bauthor{\bsnm{Wu}, \binits{J.}},
\bauthor{\bsnm{Pei}, \binits{C.}},
\bauthor{\bsnm{Lin}, \binits{X.}}, \betal:
\bctitle{Bert4rec: Sequential recommendation with bidirectional encoder representations from transformer}.
In: \bbtitle{CIKM}
(\byear{2019})
\end{bchapter}
\endbibitem

\bibitem[\protect\citeauthoryear{Li et~al.}{2022}]{RecGURU}
\begin{bchapter}
\bauthor{\bsnm{Li}, \binits{C.}},
\bauthor{\bsnm{Zhao}, \binits{M.}},
\bauthor{\bsnm{Zhang}, \binits{H.}},
\bauthor{\bsnm{Yu}, \binits{C.}},
\bauthor{\bsnm{Cheng}, \binits{L.}},
\bauthor{\bsnm{Shu}, \binits{G.}},
\bauthor{\bsnm{Kong}, \binits{B.}},
\bauthor{\bsnm{Niu}, \binits{D.}}:
\bctitle{Recguru: Adversarial learning of generalized user representations for cross-domain recommendation}.
In: \bbtitle{WSDM}
(\byear{2022})
\end{bchapter}
\endbibitem

\bibitem[\protect\citeauthoryear{Rendle}{2010}]{FM2010}
\begin{bchapter}
\bauthor{\bsnm{Rendle}, \binits{S.}}:
\bctitle{Factorization machines}.
In: \bbtitle{ICDM}
(\byear{2010})
\end{bchapter}
\endbibitem

\bibitem[\protect\citeauthoryear{Guo et~al.}{2017}]{DeepFM2017}
\begin{bchapter}
\bauthor{\bsnm{Guo}, \binits{H.}},
\bauthor{\bsnm{Tang}, \binits{R.}},
\bauthor{\bsnm{Ye}, \binits{Y.}},
\bauthor{\bsnm{Li}, \binits{Z.}},
\bauthor{\bsnm{He}, \binits{X.}}:
\bctitle{Deepfm: {A} factorization-machine based neural network for {CTR} prediction}.
In: \bbtitle{IJCAI}
(\byear{2017})
\end{bchapter}
\endbibitem

\bibitem[\protect\citeauthoryear{He et~al.}{2020}]{LightGCN2020}
\begin{bchapter}
\bauthor{\bsnm{He}, \binits{X.}},
\bauthor{\bsnm{Deng}, \binits{K.}},
\bauthor{\bsnm{Wang}, \binits{X.}},
\bauthor{\bsnm{Li}, \binits{Y.}},
\bauthor{\bsnm{Zhang}, \binits{Y.}},
\bauthor{\bsnm{Wang}, \binits{M.}}:
\bctitle{Lightgcn: Simplifying and powering graph convolution network for recommendation}.
In: \bbtitle{SIGIR}
(\byear{2020})
\end{bchapter}
\endbibitem

\end{thebibliography}


\end{document}